\documentclass[%
 reprint,
 superscriptaddress,  
 amsmath,amssymb,
 aps,
]{revtex4-2}

\usepackage{graphicx}
\usepackage{dcolumn}
\usepackage{bm}
\usepackage{amsmath,amssymb,mathtools}
\usepackage{xcolor} 
\usepackage[colorlinks=true,
            linkcolor=blue,     
            citecolor=blue,     
            urlcolor=blue]      
            {hyperref}
\usepackage[nameinlink,noabbrev]{cleveref}

\begin{document}

\preprint{APS/123-QED}

\title{Confinement-Induced Delay in Chiral Active Brownian Particles}

\author{Hrithik Barman\thanks{Email: \texttt{hrithik.udb@gmail.com}}}
\email{Email:hrithik.udb@gmail.com}
\affiliation{Indian Institute of Technology Bombay, Mumbai 400076, India}




\begin{abstract}
We investigate the stochastic dynamics of an overdamped chiral active Brownian particle confined in a harmonic potential. While delay effects and time-reversal asymmetry in active systems are commonly associated with inertial or underdamped dynamics, we show that harmonic confinement alone generates a measurable temporal lag between propulsion direction and velocity in the strictly overdamped regime. We derive exact analytical expressions for the mean-square displacement, positional cross-correlations, and a velocity–orientation delay function in closed form. Remarkably, we find that under confinement the particle’s velocity responds first to the restoring force, and the orientation follows after a finite delay. This ordering is reversed compared to inertial delay mechanisms, where velocity typically lags behind orientation due to momentum relaxation. The interplay between chirality and confinement thus produces antisymmetric temporal correlations and time-reversal asymmetry without inertial memory. Numerical simulations confirm the analytical predictions across parameter regimes. Our results establish confinement as an independent overdamped mechanism for delay and clarify the dynamical distinction between inertial and geometrically induced temporal response in active systems.
\end{abstract}

\maketitle


\section{Introduction}
Active matter systems, composed of self driven entities that dissipate energy to generate persistent motion, have emerged as a paradigm of nonequilibrium statistical physics \cite{solon2015active,basu2018active,sevilla2014theory}. Their relevance spans from biological contexts, such as bacterial swarming and the dynamics of microswimmers \cite{Lobaskin_2008,volpe2014simulation,babel2014swimming}, to the design of synthetic colloids. By constantly breaking detailed balance these systems give rise to a rich spectrum of emergent phenomena including collective motion, motility-induced phase separation \cite{Hawthorne_2025}, and non-Boltzmann stationary states featuring characteristic accumulation near confining boundaries \cite{das2018confined, di2023active}.
The interplay of Active Brownian Particles (ABPs), a canonical model in the field \cite{volpe2014simulation, majumdar2020toward}, with external confinement is a particularly fertile area of research. The harmonic trap has served as an essential tool and the resulting steady states have been a subject of intense investigation, with numerous studies exploring probability distributions, exact moments, and unique nonequilibrium properties \cite{pototsky2012active, dauchot2019dynamics, malakar2020steady, chaudhuri2021active, caraglio2022analytic, nakul2023stationary, smith2023nonequilibrium, gueneau2023active}. Beyond simple harmonic potentials confinement in other geometries such as rings \cite{Fazli_2021} or convection roll arrays \cite{Ghosh_2021}, has also been explored.
A significant extension to the ABP model is the inclusion of chirality where an intrinsic torque imposes a circular or spiral character on particle trajectories. This feature is crucial for modeling many biological and synthetic swimmers and has profound effects on both single-particle and collective behaviors. The diffusion of individual chiral particles \cite{Sevilla_2016} and their dynamics in external potentials \cite{Caprini_2023} have been established, with recent work showing how chirality can suppress phase separation \cite{Bickmann_2022}, lead to crystallization at low densities \cite{Jeong_2025}, and can be controlled or steered, as explored in recent work by Shee \cite{Shee_2025}.
Beyond chirality, particle inertia provides another crucial non-ideality. A key finding detailed extensively in the work of Löwen and collaborators is that inertia induces a non-zero delay function a lag between orientation and velocity even for free particles \cite{Scholz_2018,lowen2020inertial}. This inertial effect is distinct from other complex dynamics currently under investigation, such as the profound impact of stochastic resetting on particle search strategies and steady states \cite{shee2025resetting, Baouche_2024, Baouche_2025a}.

In this study, we investigate the combined effects of harmonic confinement and chirality on the dynamics of an active Brownian particle in the strictly overdamped regime. While inertial systems generate delay through momentum relaxation, it remains unclear how analogous temporal features arise when inertia is absent and particles are influenced instead by linear restoring forces together with chiral propulsion. To address this, we analyze the temporal relationship between the orientation angle and the instantaneous velocity direction. By explicitly comparing the evolution of these two angles, we identify a finite time lag between them and quantify it through a delay function constructed from their cross-correlation. We derive analytical expressions for key dynamical quantities, including the mean-square displacement, orientational correlations, and the delay function, and examine how its magnitude and sign depend on parameters such as trap stiffness and chirality. The theoretical predictions are validated through numerical simulations. These results demonstrate that confinement and chirality jointly produce a distinct overdamped delay mechanism, qualitatively different from inertial delay where velocity typically lags due to momentum relaxation.
\section{Model Description}
We consider a single, two-dimensional chiral Active Brownian Particle (ABP) moving with a constant self-propulsion speed $v_0$. The particle's motion is confined by an isotropic harmonic potential $U(\mathbf{r}) = \frac{1}{2} k\, r^2$ centered at the origin where $k$ is the trap stiffness. The dynamics of the particle are described by the overdamped Langevin equations\cite{dauchot2019dynamics, malakar2020steady, Caprini_2023, Pattanayak_2024} for its position $\mathbf{r}(t) = (x(t), y(t))$ and orientation angle $\phi(t)$.

The model equations are given by:
\begin{align}
\dot{\mathbf{r}}(t) &= v_0 \hat{\mathbf{n}}(t) - \mu k \mathbf{r}(t) + \sqrt{2D_t}\,\boldsymbol{\xi_t}(t) \tag{1} \label{eq:pos}\\
\dot{\phi}(t) &= \Omega + \sqrt{2D_r}\,\eta_\phi(t) \tag{2} \label{eq:angle}
\end{align}
where $\hat{\mathbf{n}}(t) = (\cos\phi(t), \sin\phi(t))$ is the instantaneous orientation of self-propulsion. The terms $\boldsymbol{\xi_t}(t)$ and $\eta_\phi(t)$ are uncorrelated Gaussian white noise sources with zero mean and delta-correlations: $\langle \boldsymbol{\xi_{i,t}}(t)\boldsymbol{\xi_{j,t}}(t) \rangle = \delta_{ij}\delta(t-t')$ and $\langle \eta_\phi(t) \eta_\phi(t') \rangle = \delta(t-t')$.

The parameter $\mu k$ has the dimension of an inverse time and defines the relaxation rate associated with the harmonic trap \cite{pototsky2012active, caraglio2022analytic, nakul2023stationary}, setting the characteristic timescale over which the particle is restored toward the origin. The constant $\Omega$ represents an intrinsic angular velocity that induces chiral motion, causing the particle to follow circular trajectories in the absence of confinement \cite{Sevilla_2016, Bickmann_2022}. The translational and rotational diffusion coefficients, $D_t$ and $D_r$, characterize thermal fluctuations in position and orientation, respectively, with rotational diffusion governing the persistence of the propulsion direction \cite{basu2018active, solon2015active}. Throughout this work, we set the mobility to $\mu = 1$ unless otherwise stated.
\begin{figure}[h]
\centering
\includegraphics[width=1\linewidth]{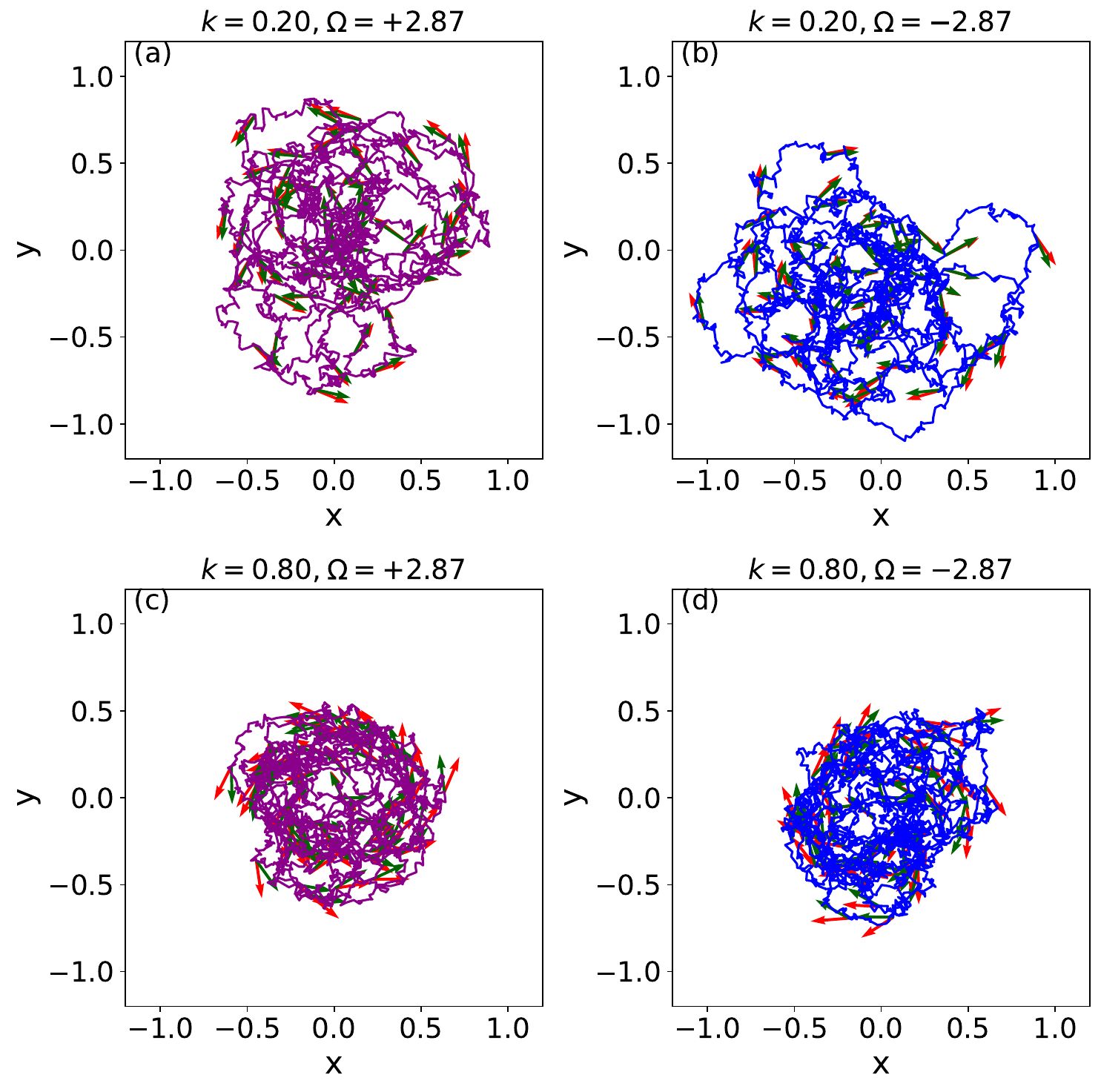}
\caption{Simulated trajectories of a chiral active Brownian particle for weak ($k=0.08$) and strong ($k=1.0$) confinement with $\Omega = \pm 2.87$. The particle trajectory is shown as a solid line, the orientation vector $\hat{\mathbf{n}}$ (red arrows), and the instantaneous velocity $\mathbf{v}$ (green arrows). Other parameters are $v_0=1.0$, $D_r=0.005$, and $D_t=0.02$.}
\label{fig:trajectories}
\end{figure}
To build geometric intuition, we examine representative trajectories of the full stochastic dynamics shown in Fig.~\ref{fig:trajectories}. The intrinsic chirality produces circular motion, while the harmonic restoring force continuously pulls the particle toward the origin, modifying these orbits. A notable feature is the systematic misalignment between the propulsion direction $\hat{\mathbf{n}}$ and the instantaneous velocity $\mathbf{v}$. In a free system these vectors coincide; however, under confinement the restoring force shifts the velocity away from the propulsion direction. This angular difference becomes more pronounced as the trap strength increases. Importantly, this misalignment does not originate from inertia, since the dynamics is strictly overdamped. Instead, it arises from the competition between chiral propulsion and linear restoring forces. As shown in the following sections, this persistent angular offset forms the geometric basis of the delay function and the associated time-reversal asymmetry.
\section{POSITION MOMENTS}

Having established the qualitative behavior of individual trajectories we now turn to a quantitative analysis of the system's statistical properties. The most fundamental of these is the ensemble average position, $\langle \mathbf{r}(t) \rangle$. This quantity represents the average trajectory that the particle would take averaged over a very large number of identical independent experiments all starting from the same initial condition.

For a simple passive particle the ensemble average position would trivially relax to the center of the trap. However for a chiral active particle the interplay between self-propulsion, chirality, and the confining force leads to more complex dynamics. We expect the average position to follow a spiral path eventually reaching a non-zero steady-state value. This final off-center position represents the balance point where the outward chiral drive is counteracted by the inward pull of the harmonic trap.

The final expressions for the x and y components of the mean position are (See Appendix\, \ref{app:moment calculation}  for details):
\begin{align}
\langle x(t)\rangle
  &= x(0)\,e^{-\mu k t}
    + \beta\, e^{-\mu k t}\Big[
      e^{\alpha t}\bigl( \alpha\cos\psi(t)+\Omega\sin\psi(t)\bigr) \notag\\
  &\qquad\qquad
      - \bigl( \alpha\cos\psi(0)+\Omega\sin\psi(0)\bigr)
    \Big] \tag{3} \label{eq:mean_x}\\[4pt]
\langle y(t)\rangle
  &= y(0)\,e^{-\mu k t}
    + \beta\, e^{-\mu k t}\Big[
      e^{\alpha t}\bigl( \alpha\sin\psi(t)-\Omega\cos\psi(t)\bigr) \notag\\
  &\qquad\qquad
      - \bigl( \alpha\sin\psi(0)-\Omega\cos\psi(0)\bigr)
    \Big] \tag{4} \label{eq:mean_y}
\end{align}
Combining these components gives the full vector expression for the radial mean position:
\begin{align}
\langle \mathbf{r}(t)\rangle
  &= \mathbf{r}(0)\,e^{-\mu k t} \notag\\
  &\quad + \beta\, e^{-\mu k t} \Bigg[
      e^{\alpha t}\vcenter{\hbox{$\displaystyle
      \begin{pmatrix}
      \alpha\cos\psi(t) + \Omega\sin\psi(t) \\[6pt]
      \alpha\sin\psi(t) - \Omega\cos\psi(t)
      \end{pmatrix}$}} \notag\\
  &\quad\;\;
      - \vcenter{\hbox{$\displaystyle
      \begin{pmatrix}
      \alpha\cos\psi(0) + \Omega\sin\psi(0) \\[6pt]
      \alpha\sin\psi(0) - \Omega\cos\psi(0)
      \end{pmatrix}$}}
    \Bigg]\tag{5} \label{eq:mean_r}
\end{align}
Where,
\[
\beta= \frac{v_0}{\alpha^2 + \Omega^2}\,  ,
\qquad \alpha= \mu k - D_r\, ,
\qquad \psi(t)= \phi_0 + \Omega t
\]\
In the absence of a harmonic trap  Eq.(\ref{eq:mean_r}) reduces to 
\begin{align}
    \langle \mathbf{r}(t)\rangle= \mathbf{r}(0)+&\frac{v_0}{D_r^2+\Omega^2}\biggl[e^{-D_r t}\vcenter{\hbox{$\displaystyle
      \begin{pmatrix}
      -D_r\cos\psi(t) + \Omega\sin\psi(t) \\[6pt]
      -D_r\sin\psi(t) - \Omega\cos\psi(t)
      \end{pmatrix}$}} \notag\\
      &+ \vcenter{\hbox{$\displaystyle
      \begin{pmatrix}
      D_r\cos\psi(0) - \Omega\sin\psi(0) \\[6pt]
      D_r\sin\psi(0) + \Omega\cos\psi(0)
      \end{pmatrix}$}}\biggr]\tag{6}
\end{align}
This result for the average position of a free active particle has been established in previous works \cite{lowen2020inertial, Pattanayak_2024}.
\begin{figure}[ht]
    \centering
    \includegraphics[width=1\linewidth]{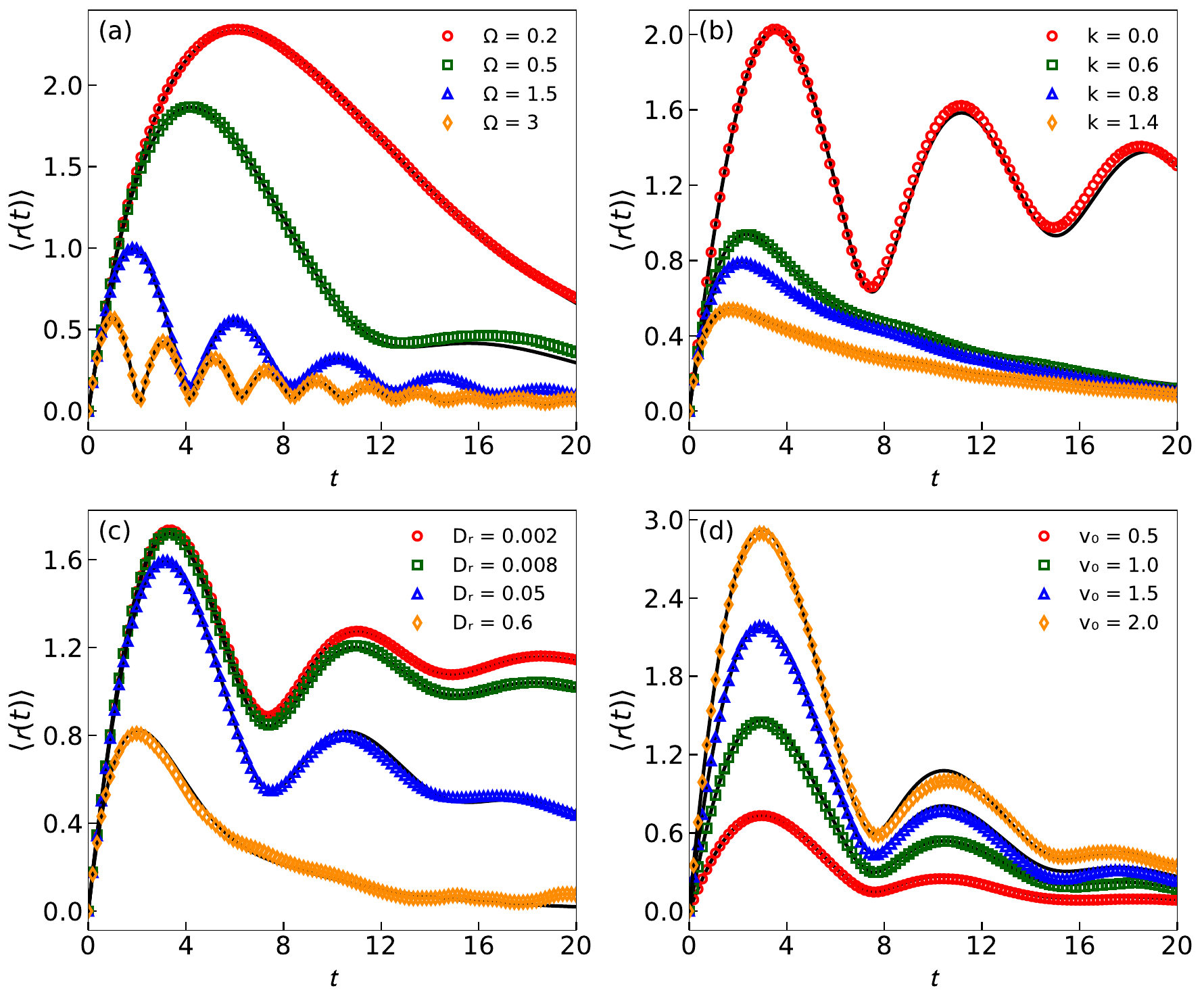} 
    \caption{Mean radial position $\langle r(t)\rangle$ of a chiral active particle in a harmonic trap. Solid lines denote theory, and symbols denote simulation obtained from the model directly. Panels show variation with (a) $\Omega$, (b) $ k$, (c) $D_r$, and (d) $v_0$.
Note: we compute the magnitude of the mean position vector, $|\langle r(t)\rangle|$ not the mean of the instantaneous magnitude, $\langle |r(t)|\rangle$.}
    \label{fig:mean position}
\end{figure}
The dynamics of the ensemble-averaged distance from the origin $\langle r(t) \rangle$, reveal how confinement, chirality, and noise shape the trajectories. At short times, $\langle r(t) \rangle$ grows rapidly as particles initially move outward in phase often overshooting their typical orbit. When the trap is present ($k>0$) the restoring force pulls particles back leading to damped oscillations around the steady state radius. In this case rotational noise $D_{r}$ controls the damping rate with stronger noise accelerating the decay of oscillations by erasing orientational memory. In contrast,\ when confinement is absent ($k=0$) the mean position does not decay but instead exhibits persistent oscillations whose frequency is set by the angular velocity $\Omega$. The propulsion speed $v_{0}$ sets the overall scale of outward excursions while $\Omega$ fixes the oscillation frequency and amplitude of the overshoot. Thus the presence or absence of confinement fundamentally changes whether $\langle r(t) \rangle$ saturates to a steady-state value or remains oscillatory.
\vspace{15
pt}\\
While the mean position reveals the average path we gain a deeper understanding of the particle's transport by analyzing the \textbf{Mean-Square  Displacement (MSD)} defined as $\langle r^2(t) \rangle$. This quantity measures how the particle spreads from its origin over time, revealing a fascinating dynamical crossover. This transition can be captured analytically and the calculation is detailed in Appendix~\ref{app:moment calculation} and the final result for a particle starting at the origin is given by the expression below. To simplify the notation we first define the following parameters:
\begin{align}
    \gamma = \mu k + D_r \, , \alpha = \mu k - D_r\, , \Delta = (\mu k)^2 - D_r^2 + \Omega^2\notag
\end{align}
With these definitions the full analytic expression for the mean squared displacement is as follows :
\begin{align}
\langle r^2(t) \rangle &= \frac{2D_t}{\mu k}(1-e^{-2\mu k t})+\frac{v_0^2\gamma}{(\gamma^2+\Omega^2)\mu k }(1-e^{-2\mu k t})\notag\\+&\frac{2v_0^2e^{-\gamma t}}{(\gamma^2+\Omega^2)(\alpha^2+\Omega^2)}\biggl[\Delta\biggl(e^{-\alpha t }-\cos(\Omega t)\biggr)\notag\\&-2D_r \Omega\sin(\Omega t)\biggr]\tag{7}\label{eq:radial MSD}
\end{align}

For very short times, we can expand the full analytic expression of the MSD. 
In the limit $t \to 0$, retaining terms up to order $t^2$, we obtain
\begin{align}
\lim_{t \to 0} \langle r^2(t) \rangle \;\approx\; 4D_t\,t \;+\; \big(v_0^2 - 4D_t\mu k\big)\,t^2 + \mathcal{O}(t^3)\tag{8}
\end{align}
\begin{figure}[ht]
    \centering
    \includegraphics[width=1\linewidth]{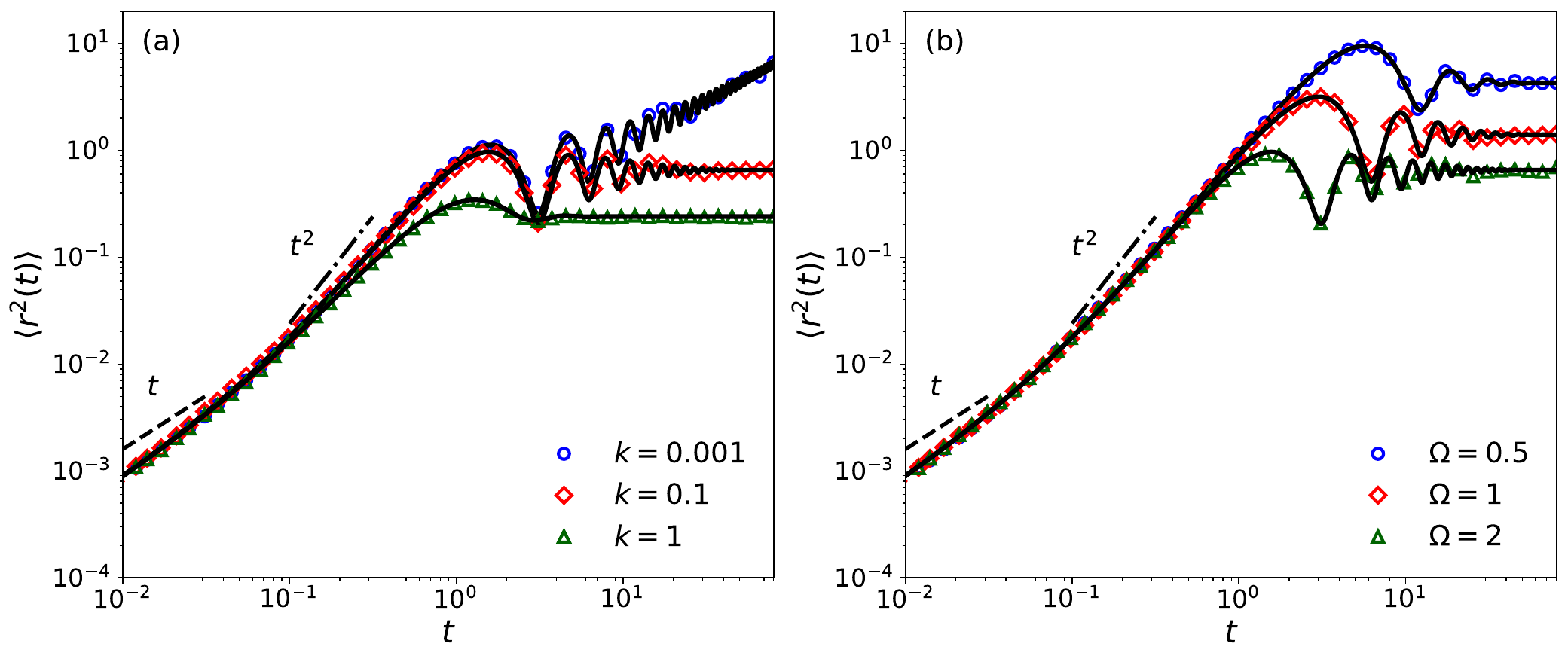} 
    \caption{
Mean square displacement $\langle r^2(t) \rangle$ for active Brownian particles with harmonic confinement and external torque. (a) Variation with confinement strength $k$ at fixed angular velocity $\Omega = 2$. (b) Variation with angular velocity $\Omega$ at fixed confinement strength $k = 0.1$. Other parameters:  $v_0 = 1$,  $D_t = 0.02$,  $D_r = 0.001$.
}
    \label{fig:msd(finite k)}
\end{figure}
The linear term $4D_t t$ corresponds to the purely diffusive contribution 
originating from translational noise. The quadratic term contains two competing effects: 
the ballistic contribution $v_0^2 t^2$ due to self-propulsion, 
and a negative correction $-\,4D_t\mu k\,t^2$ that reflects the suppression of spreading 
caused by the harmonic trap. 
Therefore, in the short-time limit the MSD grows linearly as in passive diffusion, 
with the ballistic component becoming apparent before being diminished by confinement.

In the long-time limit, the mean squared displacement saturates to
\begin{align}
\langle r^2(t \to \infty) \rangle &= \frac{2D_t}{\mu k}
+ \frac{v_0^2(\mu k + D_r)}{((\mu k + D_r)^2+\Omega^2)\mu k}\tag{9}\label{eq:msd at long time}
\end{align}

This expression represents the stationary MSD of the particle's position in the trap. 
Unlike the free active particle case where the MSD grows linearly in time and defines a long-time diffusion constant, the presence of the harmonic confinement prevents indefinite spreading. 
Instead, the MSD saturates to a finite constant value determined by the balance between active propulsion, noise, and the restoring force of the trap.

Since the MSD does not grow indefinitely, the effective long-time diffusion constant in the trap vanishes:
\begin{align}
D_L^{(\text{trap})} 
= \lim_{t \to \infty} \frac{\langle r^2(t)\rangle}{4t} = 0.
\tag{10}\label{eq:DL_trap}
\end{align}
Thus  the confined system reaches a non-equilibrium steady state characterized by a constant MSD reflecting the competition between activity and confinement.

In the absence of a confining potential ($k=0$) particle's dynamics change fundamentally. Instead of being trapped the particle is now free to diffuse indefinitely. The Mean Square displacement no longer saturates but instead grows linearly at long times defining a long-time effective diffusion coefficient. This unconfined limit provides a crucial baseline for understanding the role of confinement.

By taking the $k \to 0$ limit of Eq.~(\ref{eq:radial MSD}) and expanding the exponential terms  we obtain the expression for the MSD of a free active chiral particle\cite{lowen2020inertial, Pattanayak_2024}:
\begin{align}
\langle r^2(t) \rangle = &4D_t t + \frac{2v_0^2}{(D_r^2 + \Omega^2)^2} \bigg[ (\Omega^2 - D_r^2) + D_r(D_r^2+\Omega^2)t \notag\\&+ e^{-D_r t} \biggl( (D_r^2-\Omega^2)\cos(\Omega t) - 2D_r\Omega \sin(\Omega t) \biggr) \bigg]\tag{11}
\label{eq:msd_free_chiral}
\end{align}
\begin{figure}[ht]
    \centering
    \includegraphics[width=1\linewidth]{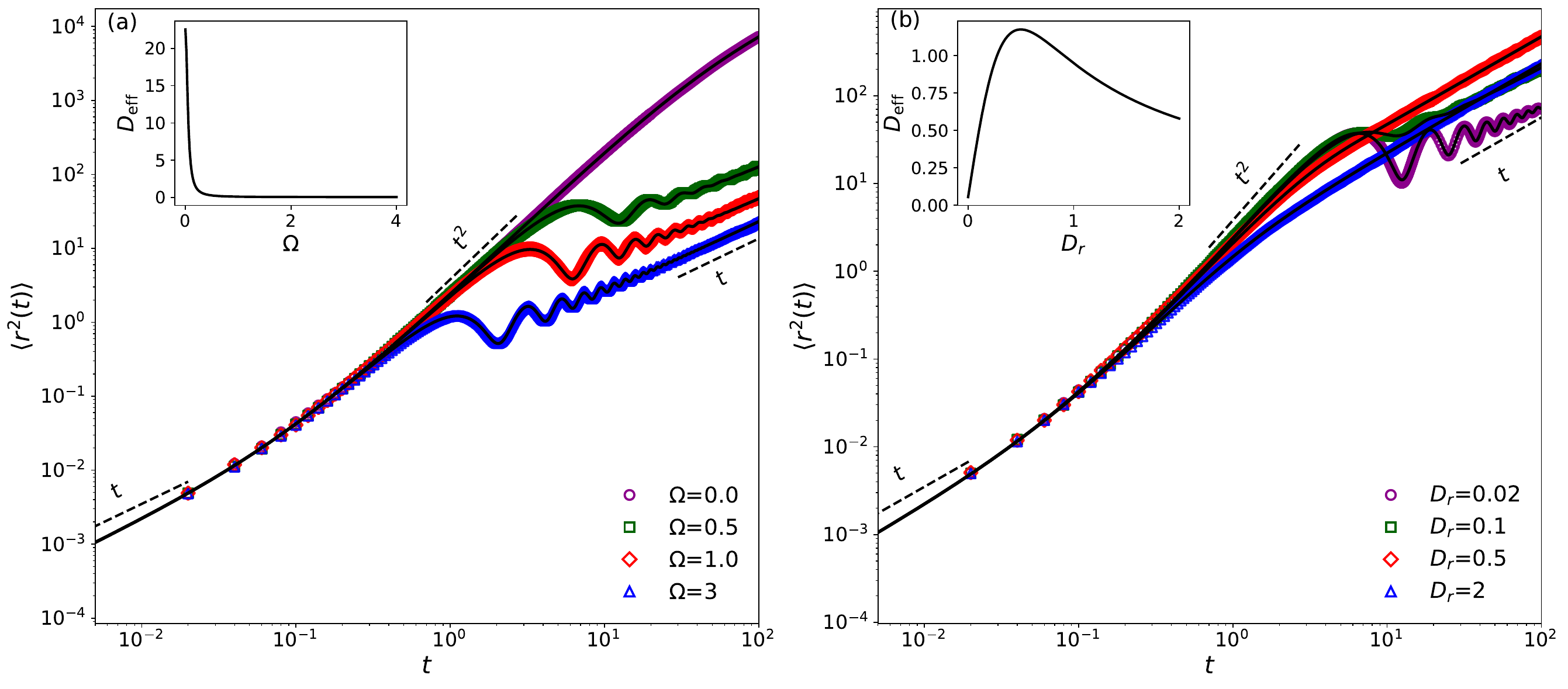} 
 \caption{Mean squared displacement (MSD) of active Brownian particles from simulations (symbols) and theoretical predictions (black lines). 
    (a) MSD for varying angular velocities $\Omega$. 
    (b) MSD for varying rotational diffusion coefficients $D_r$. 
    Insets show the corresponding effective diffusion coefficient $D_{\mathrm{eff}}$ as a function of $\Omega$ (left) and $D_r$ (right). 
    The fixed parameters are $v_0 = 1.5$ and $D_t = 0.05$.}

    \label{fig:msd_unconfined}
\end{figure}
At long times ($t \to \infty$), the exponential term vanishes, and the MSD grows linearly, $\langle r^2(t) \rangle \approx 4D_{\text{eff}}t$. By collecting all the terms proportional to $t$, we can identify the long-time effective diffusion coefficient as:
\begin{equation}
    D_{\text{eff}} = D_t + \frac{v_0^2 D_r}{2(D_r^2 + \Omega^2)}\tag{12}
    \label{eq:Deff}
\end{equation}
This important result shows how the particle's active, chiral motion enhances its ability to explore space, resulting in a diffusion rate that is significantly larger than the passive thermal diffusion, $D_t$. Fig.(\ref{fig:msd_unconfined}) shows an excellent agreement between this theory (solid lines) and our numerical simulations (markers) for the unconfined case.

Following the analysis of the mean position and mean-square displacement, we now compute the cross-correlation between the position components, $\langle x(t) y(t) \rangle$. This quantity captures how chirality couples the $x$ and $y$ motions of the particle over time. Starting from the stochastic dynamics, the cross-correlation can be formally written as an integral over the active velocity components\,(See Appendix \ref{app:moment calculation}) as :
\begin{align}
\langle x(t)y(t) \rangle 
&= v_0^2 e^{-2\mu_k t} 
\int_0^t \!\! \int_0^t e^{\mu_k (t_1 + t_2)}\notag  \\
&\quad \times 
\langle \sin \phi(t_1) \cos \phi(t_2) \rangle
\, dt_2 \, dt_1 \tag{13}
\end{align}
The angular correlation function can be evaluated using the properties of the Ornstein-Uhlenbeck process for orientation:
\begin{align}
&\langle \sin\phi(t_1)\cos\phi(t_2) \rangle =
\frac{1}{2} e^{-D_r\left|t_1-t_2\right|} \sin\Omega(t_1-t_2) \notag\\
+&\frac{1}{2} e^{-D_r(t_1+t_2+2\min(t_1,t_2))} \sin(2\phi_0 + \Omega(t_1+t_2))\tag{14}
\end{align}
Substituting this expression into the cross-correlation formula, we can write 
\begin{align}
\langle x(t)y(t) \rangle = \frac{v_0^2}{2} e^{-2\mu_k t}(I_1 +I_2)\notag \label{Eq.position cross coerelation}\tag{15}
\end{align}
where,
\begin{align}
I_1=\int_0^t \int_0^t e^{\mu_k (t_1 + t_2)}e^{-D_r\left|t_1-t_2\right|} \sin\Omega(t_1-t_2)dt_1dt_2=0\notag
\end{align}
The integral vanishes because the integrand is anti-symmetric under $t_1 \leftrightarrow t_2$ exchange while the integration domain is symmetric.

\begin{align}
I_2 
&= \int_0^t \int_0^t 
e^{\mu_k (t_1 + t_2)}
e^{-D_r(t_1 + t_2 + 2\min(t_1, t_2))} \notag\\
&\quad\times
\sin\!\big(2\phi_0 + \Omega(t_1 + t_2)\big)
\, dt_2 \, dt_1 \notag \\
&=\frac{2}{\beta^2 +\Omega^2}\biggl(\tau_1(t) -\tau_2(t)\biggr)\notag
\end{align}
where, for \(i = 1, 2\),
\begin{align}
\tau_i(t)
&= \frac{e^{s_i t}A_i(t) - A_i(0)}{s_i^2 + k_i^2}, \tag{16}
\end{align}
and
\begin{align}
A_i(t)= (\beta s_i - \Omega k_i)\sin(k_i t + 2\phi_0)\notag\\
- (\beta k_i + \Omega s_i)\cos(k_i t + 2\phi_0) \tag{17}
\end{align}

with parameters 
\begin{align}
\beta=\mu k-3D_r\,\,\,\,\,  
s_1=2\mu k-4D_r\,& \,\,\,\,
s_2=\mu k-D_r\, \notag\\
k_1=2\Omega\, \,\,\,\,
k_2=\Omega \notag
\end{align}
Therefore, we can write the full time dependent form of the cross corelation from Eq.(\ref{Eq.position cross coerelation}) as 
\begin{align}
\langle x(t)y(t) \rangle =\frac{v_0^2\,e^{-2\mu k t}}{\beta^2 +\Omega^2}\biggl(\tau_1(t) -\tau_2(t)\biggr)\tag{18}
\end{align}
\begin{figure}[h]
    \centering
    \includegraphics[width=1\linewidth]{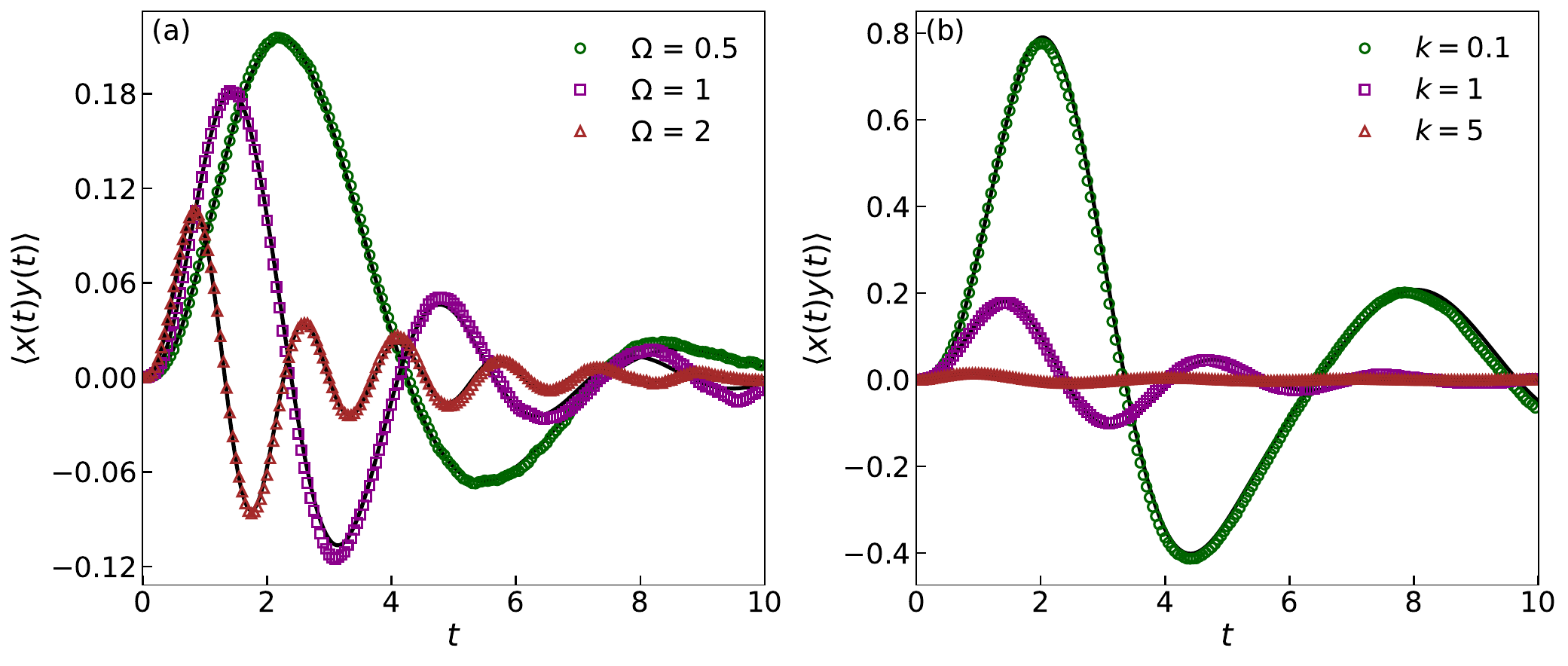}
    \caption{
Comparison of analytical (black lines) and simulated (markers) cross-correlation functions $\langle x(t)y(t) \rangle$ for a chiral active Brownian particle. 
\textbf{(a)} Effect of chirality $\Omega = 0.5\,, 1\,, 2$ at fixed trap stiffness $\kappa = 1$. 
\textbf{(b)} Effect of trap stiffness $\kappa = 0.1, 1, 5$ at fixed chirality $\Omega = 1$. 
Other parameters: $v_0 = 1$, $D_r = 0.1$, $D_t=0.01$ and $\mu = 1$.
}
    \label{fig:placeholder}
\end{figure}
The cross-correlation function $\langle x(t)y(t)\rangle$ quantifies the statistical coupling between the particle’s motion along the two spatial directions. 
For a chiral active particle this correlation exhibits oscillations that originate from the persistent rotational motion. The frequency of these oscillations is controlled by the intrinsic angular velocity $\Omega$, with larger $\Omega$ producing faster oscillatory behavior. In the presence of confinement the oscillations are no longer persistent but decay in time. This decay reflects the loss of spatial correlations due to the restoring force of the harmonic trap. 
For weak confinement the oscillatory structure remains visible over longer times, while increasing the trap stiffness $k$ accelerates the relaxation and suppresses the amplitude of the correlation. At long times $\langle x(t)y(t)\rangle$ approaches zero, indicating that the particle’s motion becomes effectively decorrelated as rotational memory and spatial excursions are damped by confinement. The combined effect of $\Omega$ and $k$ therefore determines both the frequency and the temporal persistence of the cross-correlation.
\section{Confinement Induced Delay}
As observed in the particle trajectories shown in Fig.~\ref{fig:trajectories}, 
the harmonic trap induces a clear misalignment between the particle orientation 
$\hat{\mathbf n}(t)$ and its instantaneous velocity $\dot{\mathbf r}(t)$. 
This misalignment arises from confinement. The restoring force continuously 
pulls the particle toward the trap center and prevents the velocity from 
remaining aligned with the propulsion direction.
To illustrate this dynamically, we examine the time evolution of the propulsion 
angle $\phi(t)$ and the velocity direction 
$\Theta(t)=\tan^{-1}(\dot y/\dot x)$. 
\begin{figure}[ht]
    \centering
    \includegraphics[width=1\linewidth]{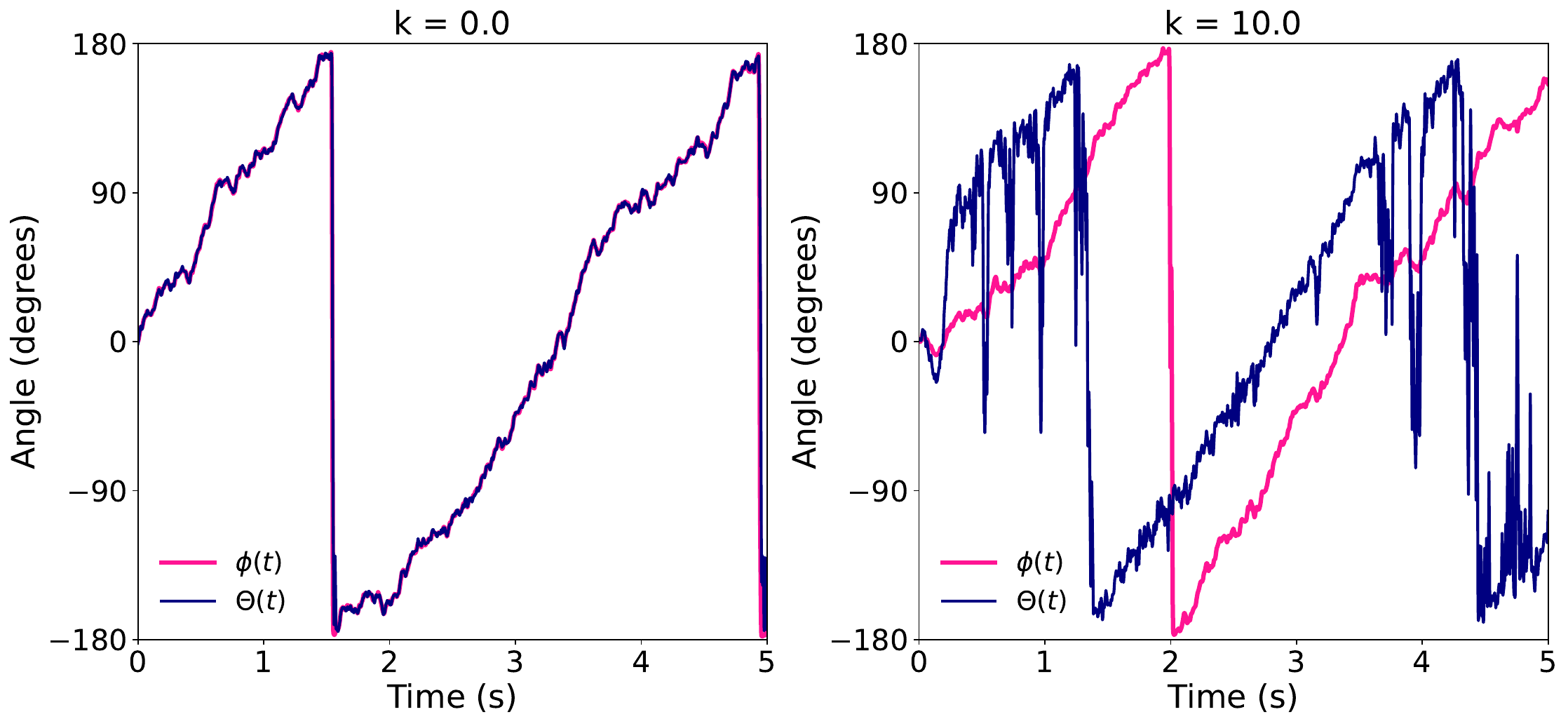}
    \caption{
Time evolution of the propulsion angle $\phi(t)$ and the velocity direction 
$\Theta(t)$. In the absence of confinement ($k=0$) the two angles overlap. 
For finite confinement ($k>0$) a phase lag appears, indicating 
confinement-induced misalignment between orientation and velocity.
}

    \label{fig:placeholder}
\end{figure}
For a free overdamped active Brownian particle ($k=0$), these two angles coincide at all times, which indicates perfect alignment between orientation and velocity. In contrast, when confinement is present ($k>0$), a persistent phase difference develops between $\phi(t)$ and $\Theta(t)$. 
This shows that the velocity does not instantaneously follow the propulsion direction. To quantify the fundamental consequences of this misalignment we can measure the breaking of time-reversal symmetry in the system.The core idea is to test if the particle's motion has a preferred arrow of time. We can check this by comparing two things: how the starting orientation affects the future velocity versus how the future orientation relates back to the starting velocity. For a simple particle in thermal equilibrium these two relationships are the same. We measure the difference between them using the function:
 $C(t) = \langle \dot{\mathbf{r}}(t).\hat{\mathbf{n}}(0) \rangle - \langle \dot{\mathbf{r}}(0).\hat{\mathbf{n}}(t) \rangle$.
A non-zero value for $C(t)$ is a clear sign that the motion is irreversible, meaning it looks wrong when played in reverse. Crucially for our model this effect is caused by the trap. For a free overdamped ABP this function is zero. Therefore a non-zero $C(t)$ directly measures the impact of confinement on the particle's dynamics.

To derive the delay function, we start from the Langevin Eq.(\ref{eq:pos})\, . 
Taking the dot product with the orientation vector $\hat{\mathbf{n}}(t)$ 
and performing ensemble averaging the resulting expressions reduce to 
correlations between position and orientation. Substituting these into 
the definition of $C(t)$ yields
\begin{align}
C(t) = -\mu k \Bigl( \langle \mathbf{r}(t).\hat{\mathbf{n}}(0)\rangle 
      - \langle \mathbf{r}(0).\hat{\mathbf{n}}(t)\rangle \Bigr) \tag{13}
\end{align}
To evaluate the simplified form we make use of the explicit solution of the Langevin equation,
\begin{align}
\mathbf{r}(t) = \mathbf{r}(0)e^{-\mu k t} 
+ v_0\int_0^t e^{-\mu k (t-s)}\hat{\mathbf n}(s)\,ds
&\notag\\+ \sqrt{2D_t}\int_0^t e^{-\mu k (t-s)}{\boldsymbol \xi_t}(s)\,ds 
\tag{14}\label{eq: position expression}
\end{align}
and set $\mathbf{r}(0)=\mathbf{0}$. The noise term vanishes upon averaging, giving
\begin{align}
\big\langle \mathbf{r}(t).\hat{\mathbf n}(0)\big\rangle
&= v_0 \int_0^t e^{-\mu k (t-s)} 
   \big\langle \hat{\mathbf n}(s).\hat{\mathbf n}(0)\big\rangle \, ds \notag\\
\big\langle \mathbf{r}(0).\hat{\mathbf n}(t)\big\rangle
&= 0  \tag{15}
\end{align}
Using the orientation correlation 
$\langle \hat{\mathbf n}(s).\hat{\mathbf n}(0)\rangle=e^{-D_r s}\cos(\Omega s)$ 
and evaluating the integral, we obtain the final expression for the delay function\,(See Appendix \ref{appendix:delay function}):
\begin{align}
C(t) = \frac{-\mu k v_0}{\alpha^2+\Omega^2}
\biggl( e^{-D_r t}\big(\alpha \cos\Omega t + \Omega \sin\Omega t\big) 
- \alpha e^{-\mu k t} \biggr)
\label{eq:delay function}\tag{16}
\end{align}
\begin{figure}[h]
    \centering
    \includegraphics[width=1\linewidth]{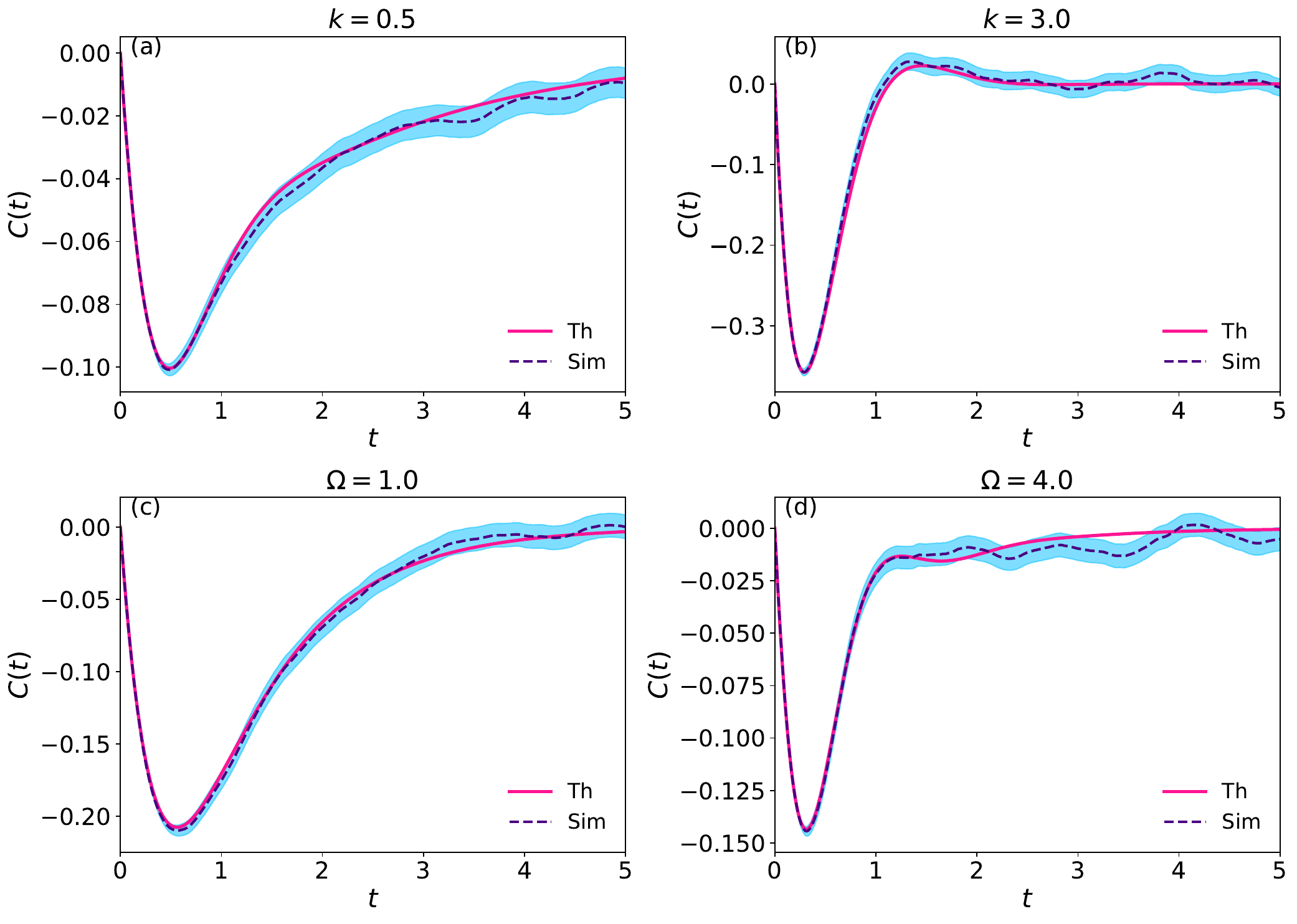}
    \caption{Delay function $C(t)$ for a confined chiral active Brownian particle.
(a,b) Variation of trap stiffness $k$ at fixed $\Omega = 2.5$.
(c,d) Variation of chirality $\Omega$ at fixed $k = 1.0$.
Solid lines are analytical results; dashed lines with shaded regions
show numerical simulations (standard error). Fixed parameters:
$v_0 = 1.0$, $D_t = 10^{-4}$.
}
    \label{fig:delay function}
\end{figure}
where $\alpha = \mu k - D_r$.\\
Fig.~(\ref{fig:delay function}) shows the time evolution of the delay 
function $C(t)$ for different system parameters, demonstrating excellent agreement between analytical theory and numerical simulations. By definition, $C(0)=0$, becomes negative at short times, reaches a minimum, and decays to zero at long times as orientational memory is lost due to rotational diffusion. The negative dip originates from confinement. As the particle begins to move, the harmonic restoring force immediately alters the velocity $\dot{\mathbf{r}}(t)$, while the orientation $\hat{\mathbf{n}}(t)$ relaxes more slowly. Thus, the velocity responds first and the orientation follows with a delay, breaking time-reversal symmetry and driving $C(t)$ negative. 
This mechanism is opposite to the inertial case, where the orientation leads and the velocity lags due to finite mass response. The depth and position of the minimum depend on system parameters. Increasing $k$ strengthens the restoring force and deepens the negative peak. Larger $D_r$ accelerates memory loss and suppresses the delay amplitude, while increasing $\Omega$ shifts the minimum to shorter times. Importantly, the observed delay arises purely from confinement in an overdamped system, establishing a distinct non-inertial route to irreversibility in active matter.
\section{Stationary Probability Distribution}
\begin{figure*}[!ht]
    \centering
    \includegraphics[width=1\textwidth]{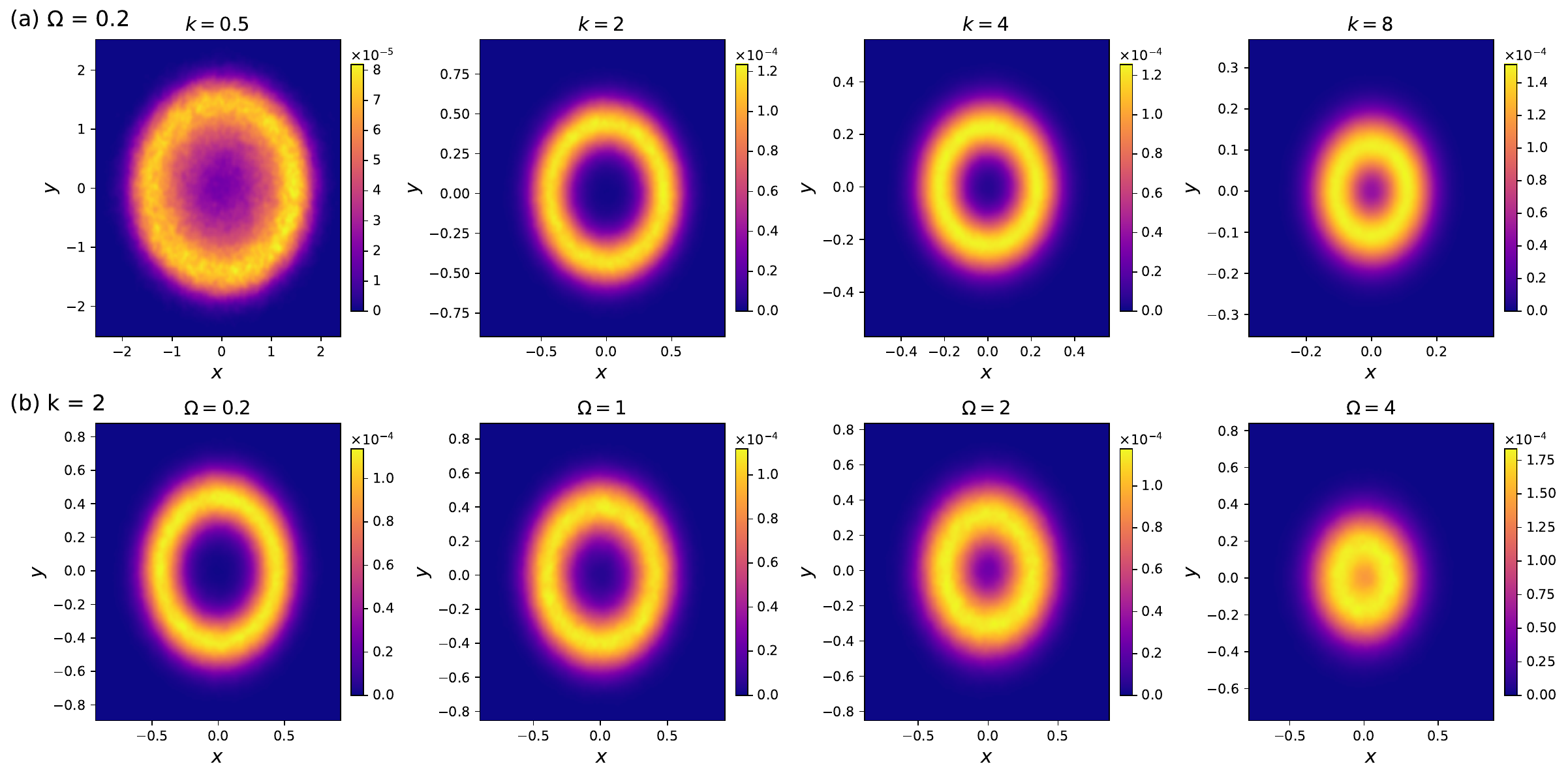} 
   \caption{ The stationary probability distribution $P(x,y)$ for a chiral ABP confined in a harmonic potential. \textbf{Top Row:} The trap strength is varied ($k=0.5, 2, 4, 8$) while the angular velocity is held fixed at $\Omega=0.2$. \textbf{Bottom Row:} The angular velocity is varied ($\Omega=0.2, 1, 2, 4$) while the trap strength is held fixed at $k=2.0$. The fixed parameters for all plots are $v_0=1.0$, $D_t=0.02$, and $D_r=0.5$. } \label{fig:prob_dist_grid_original} 
    \label{fig:prob_dist_grid} 
\end{figure*}

The stationary probability distribution of the particle's position $P(\mathbf{r})$ characterizes the system's long-term behavior. As a closed form analytical solution for the full distribution is prevented by the coupling between position and orientation we investigate its structure via direct numerical simulation of the Langevin equations (\ref{eq:pos}) and (\ref{eq:angle}). Fig.(\ref{fig:prob_dist_grid}) presents the resulting distributions for a range of trap strengths $k$ and angular velocities $\Omega$. The form of the steady state distribution is determined by the competition between confinement and activity as illustrated in Fig.(\ref{fig:prob_dist_grid}). For a fixed angular velocity $\Omega$ the trap strength $k$ sets the degree of localization. In the weak confinement limit ($k \ll 1$
) the active motion dominates and the particle explores a wide region producing a broad distribution. As $k$ increases to intermediate values the confinement compensates the spreading due to activity and the distribution becomes concentrated on a well-defined circular orbit. This regime produces the most distinct ring like pattern in the probability distribution. At strong confinement ($k \gg 1$
) the trap dominates entirely shrinking the orbit and leading to a high probability of finding the particle close to the center.  
At fixed $k$ varying the chirality $\Omega$ produces a complementary effect. For small $\Omega$ the distribution remains ring shaped but as $\Omega$ increases the circular trajectory becomes more pronounced and the stationary density takes the form of a localized ring shape with higher peaks near the center . The characteristic radius of this ring is consistent with the deterministic prediction\,(See Apendix \ref{qppendix:steady state radius})
\begin{equation}
R_{\text{st}} = \frac{v_0}{\sqrt{\Omega^2 + (\mu k)^2}} \tag{17}
\label{eq:R_st}
\end{equation}
In the limit of very large $\Omega$, the rapid reorientation effectively averages out the propulsion direction thereby enhancing localization and concentrating the distribution near the origin.
While the general chiral active case requires numerical investigation, the passive limit ($v_0 = 0$) admits an exact analytical solution. For a passive Brownian particle in a harmonic trap $U(\mathbf{r}) = \frac{1}{2}kr^2$, the overdamped Langevin equation is:

\begin{align}
\dot{\mathbf{r}}(t) = -\mu k\mathbf{r}(t) + \sqrt{2D_t}\boldsymbol{\xi_t}(t)\tag{18}
\end{align}
The corresponding Fokker-Planck equation for the probability density $P(\mathbf{r},t)$ is:
\begin{align}
\frac{\partial P(\mathbf{r},t)}{\partial t} = \nabla \cdot \left[\mu k\mathbf{r}P(\mathbf{r},t)\right] + D_t\nabla^2 P(\mathbf{r},t) \tag{19}
\end{align}
In the steady state ($\partial P/\partial t = 0$), we assume the Boltzmann form:
\begin{align}
P(\mathbf{r}) = \frac{1}{Z} \exp\left(-\frac{U(\mathbf{r})}{k_B T_{\mathrm{eff}}}\right) \tag{20}
\end{align}
With $k_B T_{\mathrm{eff}} = D_t/\mu$ and $U(\mathbf{r}) = \frac{1}{2}kr^2$, this becomes:
\begin{align}
P(\mathbf{r}) = \frac{1}{Z} \exp\left(-\frac{\mu k r^2}{2D_t}\right) \tag{21}
\end{align}
The normalization constant $Z$ is found by requiring $\int P(\mathbf{r}) d^2\mathbf{r} = 1$:
\begin{align}
\frac{1}{Z} \int_0^{2\pi} d\theta \int_0^\infty r \exp\left(-\frac{\mu k r^2}{2D_t}\right) dr = 1 \tag{22}
\end{align}

Solving the integrals yields $Z = 2\pi D_t/(\mu k)$, giving the final probability density:
\begin{figure}
    \centering
    \includegraphics[width=\columnwidth]{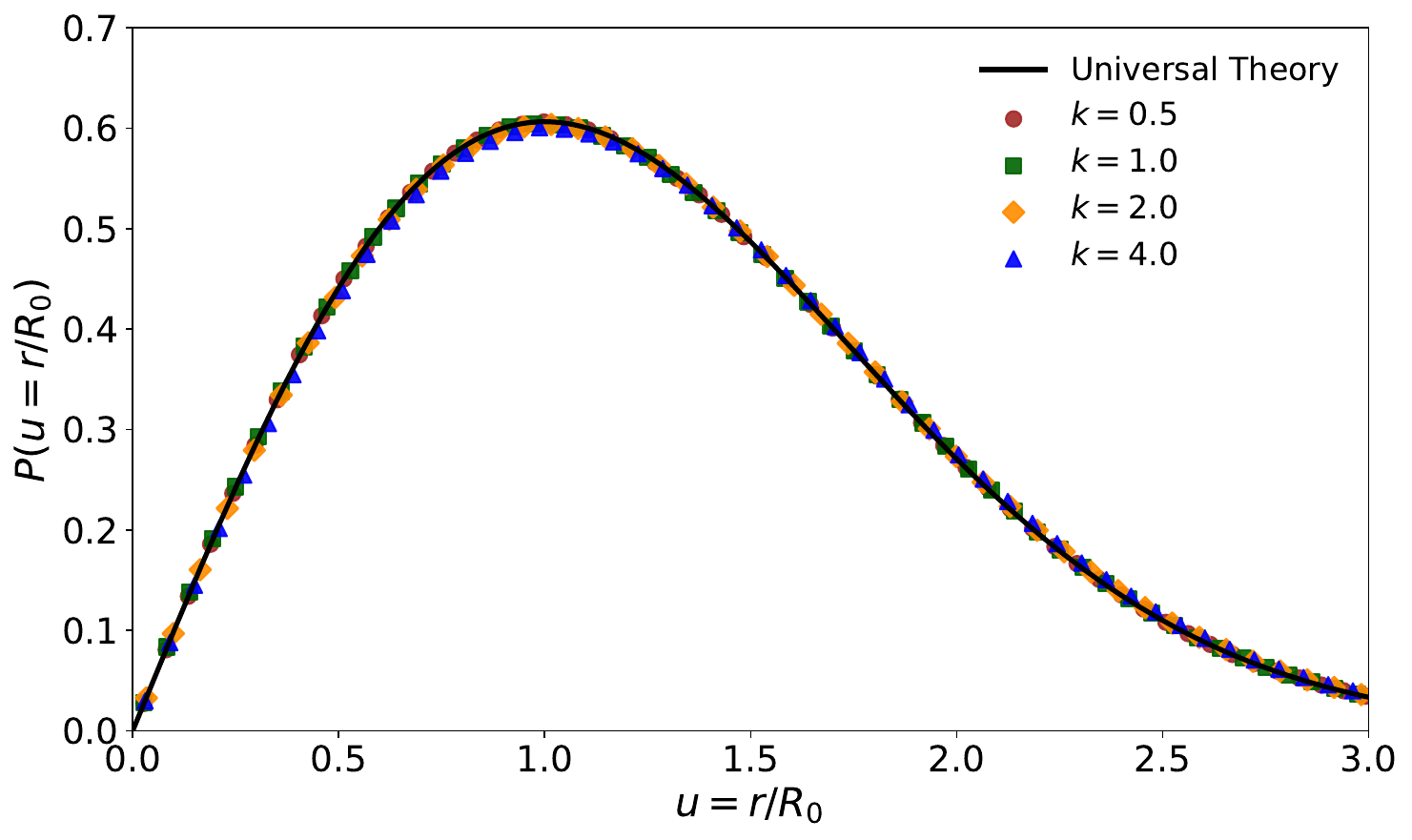}
    \caption{Universal scaling of the probability distribution for passive Brownian particles in harmonic traps. The theoretical curve (black line) shows the universal distribution $P(u) = u e^{-u^2/2}$ where $u = r/R_0$. Simulation results for different trap stiffness values $k$ (colored markers) collapse onto this universal curve when properly rescaled.}
    \label{fig:rescaled prob distribution}
\end{figure}
\begin{align}
P(\mathbf{r}) = \frac{\mu k}{2\pi D_t} \exp\left(-\frac{\mu k r^2}{2D_t}\right) \tag{23}\label{eq:prob density v0=0}
\end{align}
The corresponding radial probability distribution $p(r)$, which gives the probability density of finding the particle at a distance $r$ irrespective of direction, can be obtained by integrating $P(\mathbf{r})$ over the angular coordinate. In two dimensions this is given by
\begin{align}
p(r)\,dr = 2\pi r\, P(\mathbf{r})\, dr\tag{24}
\end{align}
Substituting the expression of $P(\mathbf{r})$ from Eq.(\ref{eq:prob density v0=0}), we obtain:
\begin{align}
p(r) = \frac{\mu k}{D_t}\, r\, \exp\!\left(-\frac{\mu k r^2}{2 D_t}\right)\tag{25}
\label{eq:pr}
\end{align}
which satisfies the normalization condition $\int_0^{\infty} p(r)\, dr = 1$.
Defining the typical length scale as 
$R_0 =(\frac{D_t}{\mu k})^{1/2}$ 
and the corresponding dimensionless radial coordinate $u = \frac{r}{R_0}$
the normalized universal form of the rescaled distribution is given by
\begin{equation}
P(u) = u\, e^{-u^2/2}\notag
\label{eq:Pu}
\end{equation}
which is independent of the specific values of $k$, $\mu$, or $D_t$. 
All rescaled distributions therefore collapse onto this single theoretical curve.

\section{CONCLUSION}
In this work we investigated the dynamics of chiral active Brownian particles confined in a harmonic potential using analytical theory together with numerical simulations. We showed that confinement changes the dynamics in a fundamental way compared to free motion. Overdamped active particles in open space maintain alignment between orientation and velocity but the introduction of a trap generates a finite delay function that acts as a clear marker of confinement induced irreversibility. We then examined the stationary probability distribution and found that the steady state evolves from broad ring shaped profiles in moderate confinement to sharply localized peaks in strong confinement. The deterministic estimate of the orbit radius provides a reliable guide for the characteristic scale of these distributions. Our results demonstrate that the combined effect of activity chirality and confinement gives rise to new nonequilibrium steady states. The analytical and numerical framework presented here can be extended to systems with many interacting particles as well as to experimental realizations of confined active matter.
\renewcommand{\appendixname}{APPENDIX}
\appendix
\section{FORMAL SOLUTION OF THE LANGEVIN EQUATION AND ORIENTATION CORRELATION}
\label{app:orientaional corelation}
This appendix provides a detailed derivation of the time-dependent solutions for the particle's position and orientation.
The particle's position vector $\mathbf{r}(t)$ is described by the following Langevin equation:
\begin{align}
   \dot{\mathbf{r}}(t)  = v_0\hat{\mathbf{n}}(t)-\mu k \mathbf{r}(t) + \sqrt{2D_t}\boldsymbol{\xi_t}(t)\label{eq:A1}
\end{align}
This is a linear, first-order inhomogeneous differential equation. We solve it using the integrating factor method. The integrating factor (IF) is:
\begin{align}
    \text{IF} = e^{\mu k t}\notag
\end{align}
Multiplying both sides of Eq.(\ref{eq:A1}) by the IF gives:
\begin{align}
    \frac{d}{dt}\left(\mathbf{r}(t)e^{\mu k t}\right) = e^{\mu k t} \left[ v_0\hat{\mathbf{n}}(t) + \sqrt{2D_t}\boldsymbol{\xi_t}(t) \right]\notag
\end{align}
We now integrate both sides with respect to a dummy time variable $s$ from the initial time $s=0$ to a final time $s=t$:
\begin{align}
    \int_0^t \frac{d}{ds}\left(\mathbf{r}(s)e^{\mu k s}\right) ds = \int_0^t e^{\mu k s} \left[ v_0\hat{\mathbf{n}}(s) + \sqrt{2D_t}\boldsymbol{\xi_t}(s) \right] ds\notag
\end{align}
Evaluating the integral gives the final formal solution for the particle's position:
\begin{align}
    \mathbf{r}(t) = \mathbf{r}(0)e^{-\mu k t} + \int_0^t e^{-\mu k(t-s)}\left[v_0\hat{\mathbf{n}}(s) + \sqrt{2D_t}\boldsymbol{\xi_t}(s)\right]ds\label{eq:A2}
\end{align}\\
The orientation angle, $\phi(t)$, evolves according to:
\begin{align}
     \dot{\mathbf{\phi}}(t)= \Omega + \sqrt{2D_r}\eta_{\phi}(t)
\end{align}
This equation can be integrated directly and therefore we can write the solution for the angle at time $t$ as :
\begin{align}
    \phi(t) = \phi_0 + \Omega t + \sqrt{2D_r}\int_0^t \eta_{\phi}(s) \,ds\label{eq:A4}
\end{align}
where $\phi_0 = \phi(0)$. 
The mean of the orientation $\phi(t)$ can be calculated as 
\begin{align}
    \langle\phi(t)\rangle = \phi_0 + \Omega t
\end{align}
To compute the second moment, $\langle\phi^2(t)\rangle$, we begin with the formal solution Eq.(\ref{eq:A4}) Squaring this expression and taking the ensemble average yields the following derivation:
\begin{align}
    \langle\phi^2(t)\rangle &= \left\langle \left( \phi_0 + \Omega t + \sqrt{2D_r}\int_0^t \eta_{\phi}(s) \,ds \right)^2 \right\rangle \nonumber \\
    &= (\phi_0 + \Omega t)^2 + 0+{2D_r}\int_0^t ds \int_0^t ds' \, \langle\eta_{\phi}(s)\eta_{\phi}(s')\rangle \nonumber \\
    &= (\phi_0 + \Omega t)^2 +{2D_r} \int_0^t ds \int_0^t ds' \,  \delta(s-s') \nonumber \\
    &= (\phi_0 + \Omega t)^2 + 2D_r \int_0^t ds \nonumber \\
    &= (\phi_0 + \Omega t)^2 + 2D_r t\label{eq:A6}
\end{align}
similarly the two point corelation can be written as 
\begin{align}
 \langle\phi(t_1)\phi(t_2)\rangle &=(\phi_0 + \Omega t_1)(\phi_0 + \Omega t_2)\notag\\&+{2D_r}\int_0^{t_1} ds \int_0^{t_2} ds' \delta(s-s')\notag \\
 &=(\phi_0 + \Omega t_1)(\phi_0 + \Omega t_2)+{2D_r}\text{min($t_1,t_2$)}\label{eq:A7} 
\end{align}
The variance and covariance can be expressed compactly given by:
\begin{align}
    \text{Var}(\phi(t)) &= \langle\phi^2(t)\rangle - \langle\phi(t)\rangle^2 = 2D_rt \notag\\
    \text{Cov}(\phi(t_1), \phi(t_2)) &= \langle\phi(t_1)\phi(t_2)\rangle - \langle\phi(t_1)\rangle\langle\phi(t_2)\rangle \nonumber \\
    & \qquad = 2D_r\min(t_1, t_2)\label{eq:A8}
\end{align}
From this we can compute the orientation correlation function, which is essential for solving the average position and mean square displacements(MSD). The orientation vector is $\hat{\mathbf{n}}(t) = (\cos\phi(t), \sin\phi(t))$. Its statistical average is found by evaluating $\langle e^{i\phi(t)} \rangle$.
\begin{align}
    \langle e^{i\phi(t)} \rangle = e^{i\langle\phi(t)\rangle} e^{-\frac{1}{2}\text{Var}(\phi(t))} = e^{i(\phi_0 + \Omega t)} e^{-D_r t}\notag
\end{align}
By using Euler's formula, $e^{i\theta} = \cos\theta + i\sin\theta$, we identify the real and imaginary parts to find the components of the mean orientation vector:
\begin{align}
    \langle\hat{\mathbf{n}}(t)\rangle =
    \begin{pmatrix}
        \langle\cos\phi(t)\rangle \\
        \langle\sin\phi(t)\rangle
    \end{pmatrix} =
    e^{-D_r t}
    \begin{pmatrix}
        \cos(\phi_0 + \Omega t) \\
        \sin(\phi_0 + \Omega t)
    \end{pmatrix}\label{eq:A9}
\end{align}
The orientation correlation function can be expressed in terms of the angle difference, $\Delta\phi = \phi(t_1) - \phi(t_2)$.
\begin{align}
    \langle\hat{\mathbf{n}}(t_1) \cdot \hat{\mathbf{n}}(t_2)\rangle &= \langle\cos(\phi(t_2) - \phi(t_1))\rangle = \langle\cos(\Delta\phi)\rangle\notag
\end{align}
For a Gaussian variable $\Delta\phi$, this average is given by $\langle\cos(\Delta\phi)\rangle = \cos(\langle\Delta\phi\rangle)e^{-\frac{1}{2}\text{Var}(\Delta\phi)}$. We first compute the mean and variance of $\Delta\phi$:
The mean is:
\begin{align}
    \langle\Delta\phi\rangle = \Omega(t_1 - t_2)
\end{align}
The variance is:
\begin{align}
    \text{Var}(\Delta\phi)=&\text{Var}(\phi(t_1))+\text{Var}(\phi(t_2))-2 \text{Cov}(\phi(t_1), \phi(t_2))\notag\\
  &=2D_r\biggl(t_1+t_2-2\min(t_1, t_2)\biggr)\notag \\&
  =2D_r|t_1-t_2|\label{eq:A11}
\end{align}
Substituting the mean and variance we can write the expression for the orientation corelation as :
\begin{align}
    \langle\hat{\mathbf{n}}(t_1).\hat{\mathbf{n}}(t_2)\rangle &= \cos(\Omega(t_1 - t_2)) \exp\left(-\frac{2D_r|t_1 - t_2|}{2}\right) \nonumber \\
    &= e^{-D_r|t_1 - t_2|} \cos\Omega(t_1 - t_2)\label{eq:A12}
\end{align}
Setting $t_1 = t$ and $t_2 = 0$ in Eq.(\ref{eq:A12}) gives
\begin{align}
   \langle \hat{\mathbf{n}}(t).\hat{\mathbf{n}}(0)\rangle
   &= e^{-D_r t}\cos(\Omega t)
   \label{eq:A13}
\end{align}

\begin{figure}[!ht]
    \centering
    \includegraphics[width=1\linewidth]{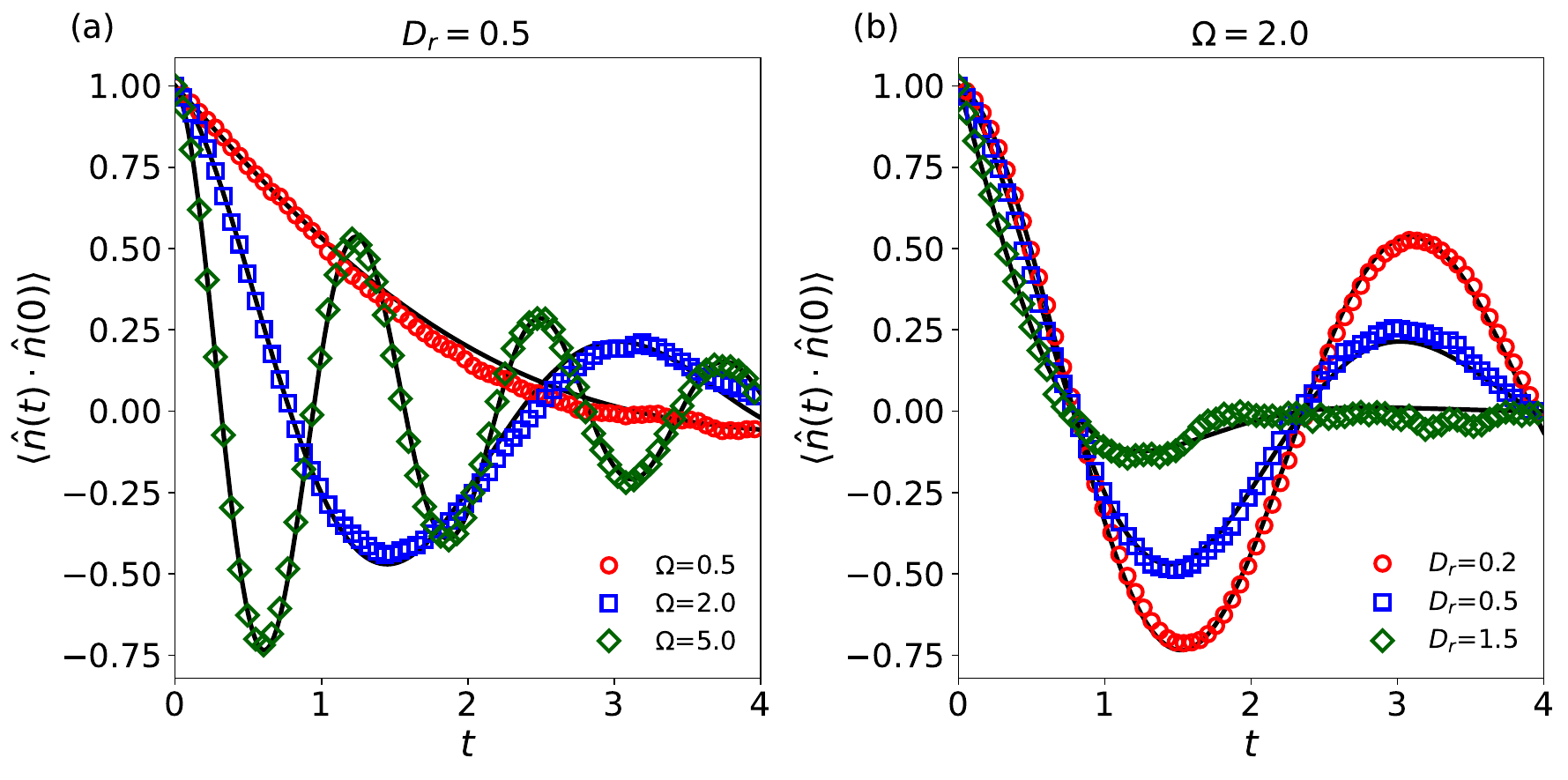} 
\caption{Comparison between theory (solid lines) and simulations (markers) for the orientation autocorrelation function. \text{Panel (a)} shows the dependence on angular velocity at fixed $D_r$, while \text{Panel (b)} shows the dependence on rotational diffusivity at fixed $\Omega$}.
    \label{fig:orientation corelation}
\end{figure}
The orientation correlation function  quantifies the memory of the particle’s initial orientation. Fig.(\ref{fig:orientation corelation}) compares the analytical prediction 
$\langle \hat{\mathbf{n}}(t).\hat{\mathbf{n}}(0)\rangle =e^{-D_r t}\cos(\Omega t)$ with numerical simulations showing excellent agreement in both cases. 
In panel~(a), we vary the angular velocity \(\Omega\) at fixed \(D_r\). 
For small \(\Omega\) the correlation decays monotonically with only weak oscillations 
dominated by the exponential damping. 
As \(\Omega\) increases, oscillatory behavior emerges clearly, reflecting the periodic 
reorientation of the particle due to deterministic rotation. 
In panel~(b), we vary the rotational diffusivity \(D_r\) at fixed \(\Omega\). 
For small \(D_r\) oscillations persist for long times whereas increasing \(D_r\) 
accelerates the decay and suppresses oscillations indicating that stochastic 
rotational noise washes out the orientation memory. 
\section{CALCULATION OF POSITION MOMENTS}
\label{app:moment calculation}
In this section we present the detailed calculation of the position moments.  
We first derive the expression for the mean position along the $x$ and $y$ directions from the Langevin equation and then compute the mean square displacement (MSD).
Starting from the formal solution of the Langevin equation\,(\ref{eq:A2})
and noting that the thermal noise has zero mean \(\langle\boldsymbol{\xi_{t}}\rangle=\mathbf0\)\, ,
the ensemble averaged position becomes
\begin{align}
\langle\mathbf{r}(t)\rangle
&= \mathbf{r}(0)\,e^{-\mu k t}
   + v_0\int_0^t e^{-\mu k (t-s)}\langle\hat{\mathbf n}(s)\rangle\,ds.
\label{eq:rmean_vector}
\end{align}
Writing \(\hat{\mathbf n}(s)=(\cos\phi(s),\sin\phi(s))\) and separating components,
\begin{align}
\langle x(t)\rangle
&=x(0)e^{-\mu k t}
   + v_0\int_0^t e^{-\mu k (t-s)}\langle\cos\phi(s)\rangle\,ds 
\label{eq:xmean_integral}\\[1pt]
\langle y(t)\rangle
&=y(0)e^{-\mu k t}
   + v_0\int_0^t e^{-\mu k (t-s)}\langle\sin\phi(s)\rangle\,ds
\label{eq:ymean_integral}
\end{align}
Using the orientation statistics Eq.(\ref{eq:A9}) we evaluate the integrals in Eqs.~\eqref{eq:xmean_integral}--\eqref{eq:ymean_integral}.
The x-component and y-component can be evaluated as
\begin{widetext}
\begin{align}
\langle x(t)\rangle &=  x(0)e^{-\mu k t} + v_0 \int_0^t e^{-\mu k (t-s)} e^{-D_r s} \cos(\phi_0 + \Omega s)  ds \notag \\
&= x(0)e^{-\mu k t} 
+ \frac{v_0 e^{-\mu k t}}{(\mu k - D_r)^2 + \Omega^2} \times \notag \\
&\quad \Biggl[ e^{(\mu k - D_r)t} \Bigl((\mu k - D_r) \cos(\phi_0 + \Omega t) \notag \\
 &+\Omega \sin(\phi_0 + \Omega t)\Bigr) 
 - \Bigl((\mu k - D_r) \cos\phi_0 + \Omega \sin\phi_0\Bigr) \Biggr]
\label{eq:B4} \\
\langle y(t)\rangle &= y(0)e^{-\mu k t} 
+ v_0 \int_0^t e^{-\mu k (t-s)} e^{-D_r s} \sin(\phi_0 + \Omega s)  ds \notag \\
&= y(0)e^{-\mu k t} 
+ \frac{v_0 e^{-\mu k t}}{(\mu k - D_r)^2 + \Omega^2} \times \notag \\
&\quad \Biggl[ e^{(\mu k - D_r)t} \Bigl((\mu k - D_r) \sin(\phi_0 + \Omega t) \notag \\
&- \Omega \cos(\phi_0 + \Omega t)\Bigr) - \Bigl((\mu k - D_r) \sin\phi_0 - \Omega \cos\phi_0\Bigr) \Biggr]
\label{eq:B5}
\end{align}

\end{widetext}
For the calculation of the mean square displacement (MSD) 
we split Eq.~(\ref{eq:A2}) into three contributions and write:
\begin{widetext}
\begin{align}
\boldsymbol{r}(t) &= \boldsymbol{r}_1(t) + \boldsymbol{r}_2(t) + \boldsymbol{r}_3(t)\, , \quad \text{with} \nonumber \\[6pt]
\boldsymbol{r}_1(t) &= \boldsymbol{r}(0)e^{-\mu k t}, \quad 
\boldsymbol{r}_2(t) = v_0 \int_{0}^{t} e^{-\mu k (t-s)}\,\hat{n}(s)\,ds, \quad
\boldsymbol{r}_3(t) = \sqrt{2D_t}\int_{0}^{t} e^{-\mu k (t-s)}\,\boldsymbol{\xi}_t(s)\,ds .
\end{align}
The mean square displacement (MSD) then follows as:
\begin{align} \langle r^2(t)\rangle &= \big\langle r_1^2(t)\big\rangle + \big\langle r_2^2(t)\big\rangle + \big\langle r_3^2(t)\big\rangle \notag \\[4pt] &\quad + 2\langle \boldsymbol{r}_1(t)\cdot\boldsymbol{r}_2(t)\rangle + 2\langle \boldsymbol{r}_1(t)\cdot\boldsymbol{r}_3(t)\rangle \notag \\&+ 2\langle \boldsymbol{r}_2(t)\cdot\boldsymbol{r}_3(t)\rangle\label{eq:B10} \end{align}
The cross terms $\langle \boldsymbol{r}_1(t)\cdot\boldsymbol{r}_3(t)\rangle\, \text{and}\,\langle \boldsymbol{r}_2(t)\cdot\boldsymbol{r}_3(t)\rangle$\, both contributes zero since $\langle\boldsymbol\xi_t\rangle=\mathbf0$ The other terms of Eq.(\ref{eq:B10}) can be calculated exactly as
\begin{align}
\text{(a)}\; \langle r_1^2(t)\rangle
&= r^2(0)\,e^{-2\mu k t} \label{eq:B11} \\[6pt]
\text{(b)}\; \langle r_3^2(t)\rangle
&= 2D_t e^{-2\mu k t} 
   \int_0^{t} ds_1 \int_0^{t} ds_2\,e^{\mu k(s_1+s_2)}\,\delta_{\alpha\beta}\,\delta(s_1-s_2) \notag \\
&= 4D_t e^{-2\mu k t} \int_0^{t} ds_1\,e^{2\mu k s_1} \notag \\
&= \frac{2D_t}{\mu k}\big(1-e^{-2\mu k t}\big)\, , 
\quad \text{where } \alpha,\beta \in \{x,y\} \label{eq:B12} \\[6pt]
\text{(c)}\; \langle r_2^2(t)\rangle
&= v_0^2 e^{-2\mu k t} 
   \int_0^t \int_0^t e^{\mu k(s_1+s_2)}\,
   \langle \hat{n}(s_1)\cdot\hat{n}(s_2)\rangle\,ds_1\,ds_2 \notag \\
&= v_0^2 e^{-2\mu k t} 
   \int_0^t \int_0^t e^{\mu k(s_1+s_2)} 
   \cos\!\big[\Omega(s_1-s_2)\big]\,e^{-D_r|s_1-s_2|}\,ds_1\,ds_2 \notag \\
&= 2v_0^2 e^{-2\mu k t} 
   \int_{s_1=0}^{t}\int_{s_2=0}^{s_1} 
   e^{\mu k(s_1+s_2)} \cos\!\big[\Omega(s_1-s_2)\big] e^{-D_r(s_1-s_2)}\,ds_2\,ds_1\, ,
   \quad \text{(taking $s_1 > s_2$)} \notag \\
&= \frac{2v_0^2 e^{-2\mu k t}}{(\mu k+D_r)^2+\Omega^2}
   \int_0^t ds_1\,e^{(\mu k-D_r)s_1}
   \Big[e^{(\mu k+D_r)s_1}(\mu k+D_r) 
   -\big((\mu k+D_r)\cos\Omega s_1 - \Omega\sin\Omega s_1\big)\Big] \notag \\
&= \frac{v_0^2 e^{-2\mu k t}}{(\mu k+D_r)^2+\Omega^2}\,
   \frac{(\mu k+D_r)}{\mu k}(1-e^{-2\mu k t}) \notag \\
&\quad + \frac{2v_0^2 e^{-2\mu k t}}{(\mu k+D_r)^2+\Omega^2}\,
   \frac{\Omega}{(\mu k-D_r)^2+\Omega^2}
   \Big[e^{(\mu k-D_r)t}\big((\mu k-D_r)\sin\Omega t - \Omega\cos\Omega t\big)+\Omega\Big] \notag \\
&\quad - \frac{2v_0^2 e^{-2\mu k t}}{(\mu k+D_r)^2+\Omega^2}\,
   \frac{(\mu k+D_r)}{(\mu k-D_r)^2+\Omega^2}
   \Big[e^{(\mu k-D_r)t}\big((\mu k-D_r)\cos\Omega t + \Omega\sin\Omega t\big)-(\mu k-D_r)\Big] \notag \\
&= \frac{v_0^2}{(\mu k+D_r)^2+\Omega^2}\,
   \frac{(\mu k+D_r)}{\mu k}(1-e^{-2\mu k t}) \notag \\
&\quad + \frac{2v_0^2 e^{-(\mu k+D_r)t}}
   {\big((\mu k+D_r)^2+\Omega^2\big)\big((\mu k-D_r)^2+\Omega^2\big)}
   \Big[\big((\mu^2 k^2+\Omega^2-D_r^2)e^{-(\mu k-D_r)t}-\cos\Omega t\big)
   -2D_r\Omega\sin\Omega t\Big] \tag{B10}
\end{align}
\begin{align}
\text{(d)}\; 2\langle \boldsymbol{r}_1(t)\cdot\boldsymbol{r}_2(t)\rangle
&=2v_0e^{-2\mu k t} \int_0^t e^{\mu k s}\,\boldsymbol{r}(0)\cdot \langle \hat{\mathbf{n}}(s)\rangle\,ds \notag \\
&=2v_0\,e^{-2\mu k t}\,\int_0^t\,e^{(\mu k -D_r)s}\,(x_0, y_0)\cdot\big(\cos(\phi_0+\Omega s),\,\sin(\phi_0+\Omega s)\big)\notag \\
&=\frac{2v_0x(0)e^{-2\mu k t}}{(\mu k -D_r)^2+\Omega^2}\biggl[e^{(\mu k -D_r)t}\biggl((\mu k-D_r)\cos(\phi_0+\Omega t)+\Omega\sin(\phi_0+\Omega t)\biggr)\notag \\
&\quad-\biggl((\mu k-D_r)\cos\phi_0+\Omega \sin\phi_0 \biggr)\biggr] \notag \\
&\quad+\frac{2v_0y(0)e^{-2\mu k t}}{(\mu k -D_r)^2+\Omega^2}\biggl[e^{(\mu k -D_r)t}\biggl((\mu k-D_r)\sin(\phi_0+\Omega t)\notag \\
&\quad-\Omega\cos(\phi_0+\Omega t)\biggr)-\biggl((\mu k-D_r)\sin\phi_0-\Omega \cos\phi_0 \biggr)\biggr]\tag{B11}
\end{align}
The final expression for the MSD is given by:
\begin{align}
\langle r^2(t)\rangle &= 
 r^2(0)\,e^{-2\mu k t}
+ \frac{2D_t}{\mu k}\big(1-e^{-2\mu k t}\big)
+ \frac{v_0^2}{(\mu k+D_r)^2+\Omega^2}\,
   \frac{(\mu k+D_r)}{\mu k}\big(1-e^{-2\mu k t}\big) \notag \\[4pt]
&\quad
+ \frac{2v_0^2 e^{-(\mu k+D_r)t}}
   {\big((\mu k+D_r)^2+\Omega^2\big)\big((\mu k-D_r)^2+\Omega^2\big)}
   \Big[\big((\mu^2 k^2+\Omega^2-D_r^2)e^{-(\mu k-D_r)t}-\cos\Omega t\big)
   -2D_r\Omega\sin\Omega t\Big] \notag \\[6pt]
&\quad
+ \frac{2v_0x(0)e^{-2\mu k t}}{(\mu k -D_r)^2+\Omega^2}\Bigg\{ 
e^{(\mu k -D_r)t}\big[(\mu k-D_r)\cos(\phi_0+\Omega t)+\Omega\sin(\phi_0+\Omega t)\big] \notag\\
&\qquad\qquad
-\big[(\mu k-D_r)\cos\phi_0+\Omega \sin\phi_0 \big]\Bigg\} \notag \\[4pt]
&\quad
+ \frac{2v_0y(0)e^{-2\mu k t}}{(\mu k -D_r)^2+\Omega^2}\Bigg\{
e^{(\mu k -D_r)t}\big[(\mu k-D_r)\sin(\phi_0+\Omega t)-\Omega\cos(\phi_0+\Omega t)\big] \notag\\
&\qquad\qquad
-\big[(\mu k-D_r)\sin\phi_0-\Omega \cos\phi_0 \big]\Bigg\}\tag{B12}
\label{eq:B15}
\end{align}
For our calculations we assume that the particle starts from the origin, 
that is, $\boldsymbol{r}(0) = (x_0,y_0) = \mathbf{0}$. \\
Now we derive the analytical expression for the cross-correlation between the particle's Cartesian position components. Starting from the formal solution of the overdamped Langevin equation with initial condition $\mathbf{r}(0) = \mathbf{0}$,
The cross-correlation function can be written as :
\begin{equation}
\langle x(t)y(t) \rangle = v_0^2 \int_0^t \int_0^t e^{-\mu k(2t-t_1-t_2)} \langle \cos\phi(t_1) \sin\phi(t_2) \rangle  dt_1 dt_2 \label{eq:xy_corr_start}\tag{B13}
\end{equation}
where the noise terms average to zero due to their uncorrelated nature: $\langle \xi_{t,x}(t_1)\xi_{t,y}(t_2) \rangle = 0$.
We employ the trigonometric identity:
\begin{align}
\sin\phi(t_1)\cos\phi(t_2) = \frac{1}{2}\sin(\phi(t_1) - \phi(t_2)) + \frac{1}{2}\sin(\phi(t_1) + \phi(t_2)) \label{eq:trig_identity}\tag{B14}
\end{align}
Using the identity $\langle \sin(X) \rangle = \sin(\langle X \rangle)e^{-\frac{1}{2}\text{Var}(X)}$ for Gaussian $X$, we obtain:
\begin{align}
\langle \sin\phi(t_1)\cos\phi(t_2) \rangle = \frac{1}{2}e^{-D_r|t_1-t_2|}\sin[\Omega(t_1-t_2)] + \frac{1}{2}e^{-D_r(t_1+t_2+2\min(t_1,t_2))}\sin[2\phi_0 + \Omega(t_1+t_2)] \label{eq:angular_corr_full}\tag{B15}
\end{align}
Substituting equation (\ref{eq:angular_corr_full}) back into equation (\ref{eq:xy_corr_start}) yields:
\begin{align}
\langle x(t)y(t) \rangle = \frac{v_0^2}{2}e^{-2\mu k t}(I_1 + I_2) \label{eq:xy_corr_decomp}\tag{B16}
\end{align}
Where\, ,
\begin{align}
I_1 &= \int_0^t \int_0^t e^{\mu k(t_1+t_2)} e^{-D_r|t_1-t_2|} \sin[\Omega(t_1-t_2)] dt_1 dt_2 \label{eq:I1_def}=0\tag{B17}
\end{align}
\begin{align}
I_2 &= \int_0^t \int_0^t e^{\mu k(t_1+t_2)} e^{-D_r(t_1+t_2+2\min(t_1,t_2))} \sin[2\phi_0 + \Omega(t_1+t_2)] dt_1 dt_2 \notag\\
&= 2\int_{t_1=0}^t e^{(\mu k -D_r)t_1}dt_1 \int_{t_2=0}^{t_1}e^{(\mu k -3D_r)t_2 }\,\sin[2\phi_0 + \Omega(t_1+t_2)] dt_2 \notag\\
&= 2\int_{t_1=0}^t e^{(\mu k -D_r)t_1} \bigg[ \frac{e^{(\mu k-3D_r)t_2}}{(\mu k-3D_r)^2 + \Omega^2} \biggl( (\mu k-3D_r)\sin(2\phi_0 + \Omega(t_1+t_2)) - \Omega\cos(2\phi_0 + \Omega(t_1+t_2))\biggr)\biggl]_{t_2=0}^{t_1} dt_1 \notag\\
&=\frac{2}{\beta^2 +\Omega^2}\biggl(\tau_1(t) -\tau_2(t)\biggr)\tag{B18}
\end{align}
Thus the final simplified expression of the cross corelation is given by
\begin{align}
\langle x(t)y(t) \rangle =\frac{v_0^2\,e^{-2\mu_k t}}{\beta^2 +\Omega^2}\biggl(\tau_1(t) -\tau_2(t)\biggr)\tag{B19} 
\end{align}
\end{widetext}

\begin{widetext}
\section{DERIVATION OF DELAY FUNCTION: }
\label{appendix:delay function}
In this appendix we derive the delay function
\begin{equation}
C(t) = \langle \dot{\mathbf r}(t)\cdot\hat{\mathbf{n}}(0)\rangle - \langle \dot{\mathbf r}(0)\cdot\hat{\mathbf{n}}(t)\rangle\notag
\end{equation}

The Langevin equations are
\begin{align}
\dot{\mathbf r}(t) &= v_0 \hat{\mathbf{n}}(t) - \mu k\, \mathbf r(t) + \sqrt{2D_t}\,\boldsymbol{\xi_t}(t) \notag\\
\dot\phi(t) &= \Omega + \sqrt{2D_r}\,\eta_\phi(t)\,, \qquad \hat n(t) = (\cos\phi(t)\notag \sin\phi(t))
\end{align}
Dotting the position equation with $\hat{\mathbf{n}}(0)$ and averaging gives
\begin{align}
\langle \dot{\mathbf r}(t)\cdot \hat{\mathbf{n}}(0)\rangle
= v_0 \langle \hat{\mathbf{n}}(t)\cdot \hat n(0)\rangle
- \mu k \langle \mathbf r(t)\cdot \hat{\mathbf{n}}(0)\rangle\label{eq:C1}
\end{align}
Similarly at $t=0$\, ,

\begin{align}
\langle \dot{\mathbf r}(0)\cdot \hat{\mathbf{n}}(t)\rangle
= v_0 \langle\hat{\mathbf{n}}(0)\cdot \hat{\mathbf{n}}(t)\rangle
- \mu k \langle \mathbf r_0\cdot\hat{\mathbf{n}}(t)\rangle\label{eq:C2}
\end{align}
Subtracting, and using $\langle \hat{\mathbf{n}}(t)\cdot\hat{\mathbf{n}}(0)\rangle = \langle \hat{\mathbf{n}}(0)\cdot \hat{\mathbf{n}}(t)\rangle$, we obtain
\begin{align}
C(t) = -\mu k \biggl( \langle \mathbf r(t)\cdot \hat{\mathbf{n}}(0)\rangle - \langle \mathbf r_0\cdot \hat{\mathbf{n}}(t)\rangle \biggr)\label{eq:C3}
\end{align}
The formal solution of $\mathbf r(t)$ is
\begin{align}
\mathbf r(t) = \mathbf r_0 e^{-\mu k t} + v_0 \int_0^t e^{-\mu k (t-s)} \hat{\mathbf{n}}(s)\,ds
+ \sqrt{2D_t}\int_0^t e^{-\mu k (t-s)} \boldsymbol\xi_t(s)\,ds \notag
\end{align}
Averaging, and dotting with $\hat{\mathbf{n}}(0)$, gives
\begin{align}
\langle \mathbf r(t)\cdot \hat{\mathbf{n}}(0)\rangle
= e^{-\mu k t}\langle \mathbf r_0\cdot \hat{\mathbf{n}}(0)\rangle
+ v_0 \int_0^t e^{-\mu k (t-s)} \langle \hat{\mathbf{n}}(s)\cdot \hat{\mathbf{n}}(0)\rangle\,ds\label{eq:C4}
\end{align}
The orientation correlation is
\begin{align}
\langle \hat{\mathbf{n}}(s)\cdot \hat{\mathbf{n}}(0)\rangle = e^{-D_r s}\cos\Omega s\notag
\end{align}
Thus
\begin{align}
\langle \mathbf r(t)\cdot \hat{\mathbf{n}}(0)\rangle
= &e^{-\mu k t}\langle \mathbf r_0\cdot \hat{\mathbf{n}}(0)\rangle
+ v_0 \int_0^t e^{-\mu k (t-s)} e^{-D_r s}\cos(\Omega s)\,ds\notag \\
&=e^{-\mu k t}\biggl[x_0\cos \phi_0+y_0\sin \phi_0\biggr]+ \frac{v_0\,e^{-\mu k t}}{(\mu k -D_r)^2 \,+\Omega^2}\biggl[e^{(\mu k -D_r)t}\biggl((\mu k-D_r)\cos \Omega t\,+\Omega\sin\Omega t\biggr)-(\mu k -D_r)
\biggr]\label{eq:C5}
\end{align}
similarly second term of Eq.(\ref{eq:C3}) gives
\begin{align}
   \langle \mathbf r_0\cdot \hat{\mathbf{n}}(t)\rangle=&\mathbf r_0\cdot\langle\hat{\mathbf{n}}(t)\rangle\notag\\&=x_0e^{-D_r t}\cos(\phi_0 +\Omega t)\,+y_0e^{-D_r t}\sin(\phi_0 +\Omega t)\label{eq:C6}
\end{align}
Putting these in Eq.(\ref{eq:C3}) we get the final expression for the delay function as :
\begin{align}
C(t) = &\mu k\,e^{-D_r t}\biggl[x_0e^{-D_r t}\cos(\phi_0 +\Omega t)\,+y_0e^{-D_r t}\sin(\phi_0 +\Omega t)\biggr]-\mu k e^{-\mu k t}\biggl[x_0\cos \phi_0+y_0\sin \phi_0\biggr] \notag\\
&+\frac{v_0 \mu k \,e^{-\mu k t}}{(\mu k -D_r)^2 \,+\Omega^2}\biggl[(\mu k -D_r)\,-\,e^{(\mu k -D_r)t}\biggl((\mu k-D_r)\cos \Omega t\,+\Omega\sin\Omega t\biggr)
\biggr]\label{eq:C7}
\end{align}
For the central case with initial condition \(\mathbf{r}_0 = 0\), the delay function simplifies significantly. The resulting expression, Eq.~(\ref{fig:delay function}), is the key analytical result used throughout our analysis and provides the basis for comparison with numerical simulations. It captures the system's delayed response driven solely by the initial self-propulsion velocity \(v_0\), through the interplay of relaxation \(\mu k\), rotational diffusion \(D_r\), and external driving \(\Omega\).
\end{widetext}
\section{STEADY STATE CIRCULATION RADIUS}
\label{qppendix:steady state radius}
To derive the steady-state circulation radius $R_{\textbf{st}}$ for this model, we employ a complex number representation of its equations of motion. This approach mathematically offers a more compact and elegant solution than handling the Cartesian components separately, as it naturally encapsulates the rotational symmetry of the problem. The complex formulation reduces the system of coupled equations to a single first-order linear differential equation, which can be solved exactly.
The noise free equations are given by   
\begin{align}
    \dot{x}(t) &= -\mu k x(t) + v_0 \cos\phi(t)\notag \\
    \dot{y}(t) &= -\mu k y(t) + v_0 \sin\phi\notag\\
    \dot{\phi}(t) &= \Omega \notag
\end{align}
Introducing the complex position $z(t) = x(t) + i y(t)$ and the orientation vector $\hat{\mathbf{n}}(t) = e^{i\phi(t)}$, the equations of motion reduce to a compact form:  
\begin{align}
    \dot{z}(t) &= -\mu k\, z(t) + v_0 e^{i\phi(t)}, \label{eq:D1} \\
    \dot{\phi}(t) &= \Omega. \label{eq:D2}
\end{align}
From Eq.(\ref{eq:D2}), the orientation evolves as:
\begin{align}
    \phi(t) = \Omega t + \phi_0.
\end{align}
The solution od Eq.(\ref{eq:D1}) is given by:
\begin{align}
    z(t) 
    &= z(0) e^{-\mu k t} 
    + \frac{v_0 e^{i\phi_0}}{\mu k + i\Omega} \left( e^{i\Omega t} - e^{-\mu k t} \right). 
    \label{eq:full_solution}
\end{align}
In the long-time limit $t \gg \tau_k$, the transient contributions vanish on the trap relaxation timescale $\tau_k = 1/(\mu k)$, leaving only the steady circular motion.

\begin{align}
    z_{\text{st}}(t) 
    &= \frac{v_0 e^{i\phi_0}}{\mu k + i\Omega}\, e^{i\Omega t}.
\end{align}
This represents circular motion of radius
\begin{align}
    R_{\text{st}} 
    &= \left| z_{\text{st}}(t) \right| 
     = \frac{v_0}{\sqrt{(\mu k)^2 + \Omega^2}}, 
    \label{eq:D6}
\end{align}

\nocite{*}
\bibliography{apssamp}

\end{document}